\documentclass[twocolumn]{aastex63}
\usepackage{amsmath}
\usepackage{mathtools}
\usepackage{xcolor}

\def\Ha{{\rm H}$\alpha$}
\def\Hb{{\rm H}$\beta$}
\def\O3{{\rm O}III}
\def\N2{{\rm N}II}
\def\S2{{\rm S}II}

\shorttitle{Spectra Decomposition}
\shortauthors{Teimoorinia et al.}

\graphicspath{{./}{figures/}}
\usepackage{wrapfig}

\begin{document}

\title{
Revisiting AGN Placement on the BPT Diagram: A Spectral Decomposition Approach
}

\correspondingauthor{Hossen Teimoorinia}
\email{hossen.teimoorinia@nrc-cnrc.gc.ca, hossteim@uvic.ca}

\author{Teimoorinia, Hossen}
\affiliation{NRC Herzberg Astronomy and Astrophysics, 5071 West Saanich Road, Victoria, BC, V9E 2E7, Canada}
\affiliation{Department of Physics and Astronomy, University of Victoria, Victoria, BC, V8P 5C2, Canada}

\author{Shishehchi, Sara}
\affiliation{Signify, Research US, 1 Van de Graaff Drive, Burlington, MA 01803, USA}

\author{Archinuk, Finn}
\affiliation{Department of Biomedical Engineering, University of Alberta, Edmonton AB, T6G 1H9, Canada.}

 \author{Woo, Joanna}
\affiliation{Department of Physics, Simon Fraser University, 8888 University Dr, Burnaby BC, V5A 1S6, Canada}

\author{Bickley, Robert}
\affiliation{Department of Physics and Astronomy, University of Victoria, Victoria, BC, V8P 5C2, Canada}

\author{Lin, Ping}
\affiliation{Ciena, 5050 Innovation Drive, Ottawa, ON, K2K 0J2, Canada}

\author{Hu, Zhonglin}
\affiliation{NRC Herzberg Astronomy and Astrophysics, 5071 West Saanich Road, Victoria, BC, V9E 2E7, Canada}
\affiliation{Department of Electrical and Computer Engineering, University of Victoria, 3800 Finnerty Road, Victoria BC, V8P 5C2, Canada}

\author{Petit, Emile}
\affiliation{Departement of Mathematics and Statistics, McGill University, 845 Sherbrooke Street West, Montreal, QC, H3A 0G4, Canada}

\begin{abstract}

Traditional single-fibre spectroscopy provides a single galaxy spectrum, forming the basis for crucial parameter estimation. However, its accuracy can be compromised by various sources of contamination, such as the prominent \Ha~emission line originating from both Star-Forming (SF) regions and non-Star-Forming regions (NonSF), including  Active Galactic Nuclei (AGN). The potential to dissect a spectrum into its SF and NonSF constituents holds the promise of significantly enhancing precision in parameter estimates. In contrast, Integral Field Unit (IFU) surveys present a solution to minimize contamination. These surveys examine spatially localized regions within galaxies, reducing the impact of mixed sources. Although an IFU survey's resulting spectrum covers a smaller region of a galaxy than single-fibre spectroscopy,  it can still encompass a blend of heterogeneous sources.  Our study introduces an innovative model informed by insights from the MaNGA IFU survey. This model enables the decomposition of galaxy spectra, including those from the Sloan Digital Sky Survey (SDSS), into SF and NonSF components. Applying our model to these survey datasets produces two distinct spectra, one for SF and another for NonSF components, while conserving flux across wavelength bins.  When these decomposed spectra are visualized on a BPT diagram, interesting patterns emerge. There is a significant shift in the placement of the NonSF decomposed spectra, as well as the emergence of two distinct clusters in the LINER and Seyfert regions. This shift highlights the key role of SF `contamination' in influencing the positioning of NonSF spectra within the BPT diagram.

\end{abstract}

\keywords{Galaxies - galaxies: abundances - galaxies: star formation- galaxies: active- methods: data analysis - galaxies: nuclei- galaxies: statistics} 

\section{Introduction}
\label{sec:introduction}

Galaxies are complex systems that require systematic classification to enable thorough analysis. These classifications include distinguishing between star-forming and non-star-forming (NonSF) activities, including AGNs, which are potent energy sources located at the centers of galaxies. They emit energy across a broad electromagnetic spectrum, from radio waves to X-rays \citep[e.g.,][]{Urry95, Rubinur24,vez24}. NonSF galaxies are further subdivided into two categories \citep{Kewley06, Cid10} of Seyfert \citep[e.g.,][]{Winter11, Ackermann12, Yao23} and Low Ionization Nuclear Emission-Line Region (LINER) \citep[e.g.,][]{Heckman80, Kauffmann09, Yan12, Singh13, Scott14, Herpich16, Coldwell17, Coldwell18, Houston23} galaxies. In these galaxies, hot and old stars play a significant role in producing low-ionization emissions. They are older, more massive, less dusty, less concentrated, and have higher velocity dispersions compared to Seyferts \citep{Kewley06}. The emission from these galaxies can dominate the overall galactic emission, especially at certain wavelengths. This dominance is critical as NonSF emissions can obscure the wavelengths important for identifying star-forming activities, particularly in the ultraviolet and infrared spectra. These wavelengths are vital for accurately estimating the Star Formation Rate (SFR) \citep[e.g.,][]{Tacchella22}. Given that the spectra under study combine emissions from HII regions, pure Seyfert and LINER AGNs, and LINER-like emissions resulting from shocks or photoionization by old, UV-bright stars, we may use the terminology 'SF' to refer to star-forming regions and 'NonSF' for non-star-forming regions, which may also include true AGNs, in the context of our paper.

Overall, these decomposition distributions show a significant narrowing of AGN density.  Contamination of AGN spectra is well discussed in the literature, with a notable example being low ionization nuclear Emission-Line Region (LINER) galaxies. These galaxies are known for their diverse properties, which have been the subject of extensive studies (e.g., \cite{Scott14}; \cite{Herpich16}; \cite{Coldwell17}; \cite{Coldwell18}, \cite{Houston23}). We refer readers to \cite{Coldwell18}, who provides evidence supporting the role of hot and old stars in generating low-ionization emissions in these galaxies, which can be considered a sort of contamination.

Galactic research utilizes diverse data acquisition methodologies, such as photometric data collection and integrated single-fibre spectroscopy. The Sloan Digital Sky Survey (SDSS) (\cite{York2000a}; \cite{Strauss2002}) employs a spectrograph to capture galactic spectra. These spectra may not encompass the entire galaxy, highlighting the importance of the covering fraction. The spectrum that emerges is a composite of emissions from various parts of the galaxy, each rich in emission lines necessary for classifying galaxies based on their physical properties. These complex optical spectra can be dissected to isolate contributions from  SF and NonSF activities, as demonstrated by \cite{Davies17}. The research delves into the challenging task of differentiating emissions caused by star formation from those arising from NonSF activities. Employing the superposition principle \citep[e.g., ][ for a general review of the principle]{Gunther2019, Gan2022}, the study reveals that emissions from NonSF and star-forming regions within a galaxy can be viewed as distinct but overlapping waveforms, together forming the galaxy's overall emission spectrum.

The superposition principle asserts that any resultant waveform at a given point, created by the overlap of multiple waves, is simply the algebraic summation of these individual waves. Consequently, the observed emission from a galaxy often represents a blend of contributions from both NonSF and SF regions. Accurately distinguishing these emissions is crucial for precise astronomical observations and interpretations due to their different mechanisms and inherent properties \citep{StasinskaBook}. Regarding the application of the superposition principle, modelling a galaxy's spectrum becomes a complex task due to uncertainties and complexities related to star composition and formation. To streamline this process, researchers often utilize a linear superposition of `Simple Stellar Populations' (SSPs), which treats galaxies as assemblies of stars that share similar ages and chemical compositions. For example,  \cite{Bruzual03} illustrates this approach by reconstructing a galaxy's integrated spectrum through a linear combination of individual stellar spectra from various types drawn from an extensive and diverse spectral library to ensure a broad representation of stellar characteristics. Additionally, the stellar initial mass function plays an important role in the formation of SSPs \citep[e.g.,][]{Salpeter55}.

Regarding the above example, the spectrum of a galaxy, represented by N data points, can be viewed as a combination of SSPs, each weighted by an SFR. With a given integrated spectrum, the goal can be to identify the contributing SSPs (each with N data points) along with their respective weights (i.e., the SFRs). These kinds of tasks are analytically complex, as they involve solving numerous equations to determine the N data points of SSPs with the appropriate SFR weights. Alternatively, these kinds of challenging problems may be addressed using machine learning by providing a robust training set of integrated spectra and the contributing SSPs. \cite{Woo2024} used a deep learning approach to estimate mass-weighted stellar age and metallicities of the SSPs that comprise galaxy spectra.  A deep neural network can also be employed to adjust the model's parameters to find the reverse process. In this paper, we propose a method to reverse the procedure by using a data set that includes `pure' SF and NonSF galaxy spectra.  We will demonstrate how to decompose a composite spectrum (with N data points) into two components: NonSF and SF, each with the same number of data points and corresponding weights.

To classify NonSF and SF galaxies, Baldwin, Phillips, and Terlevich (BPT; \cite{baldwin1981}) introduce a classification system using emission line ratios. For galaxy classification via BPT diagrams, the theoretical and empirical demarcation lines are introduced by Kewley et al. (2001; K01) and Kauffmann et al. (2003; K03). K01 delineated the NonSF-dominated region, while K03 defined the SF galaxy region. These lines were applied to SDSS galaxies, and those situated between these two lines were categorized as composites. The BPT diagram, which uses log [OIII]/H$_\beta$ vs. log [NII]/H$_\alpha$ for classification, is not the only method to classify NonSF and SF galaxies. For example, the WHAN method (Cid-Fernandes et al., 2010, 2011) uses log EW$_{H\alpha}$ vs. log [NII]/H$_\alpha$ and classifies galaxies into four different classes: `strong AGNs,' `weak AGNs,' SF, and passive galaxies. The effectiveness of equivalent width emission lines in classifying SF and NonSF galaxies is also demonstrated by \cite{Teimoorinia-2018}, who used the SDSS (DR7) dataset and supervised Deep Neural Networks (DNN). \cite{Souza17} employ a machine learning approach, specifically a Gaussian mixture model (an unsupervised method), along with SDSS (DR7) and SEAGal/STARLIGHT, to analyze the BPT and WHAN diagrams. Their best-fit model identifies four Gaussian components that can account for up to 97\% of the data variance. However, their analysis does not provide statistical evidence supporting the presence of a Seyfert/LINER dichotomy within their sample.

While SDSS, an example of a single-fibre spectroscopic survey, has significantly contributed to our understanding of galaxy formation and evolution, it is limited by averaging the complex internal structures of target galaxies. For a more detailed study, \cite{Davies16} use data from the S7 survey \citep[see ][for the Integral Field Unit (IFU) survey and data]{Dopita15} on NGC 5728 and NGC 7679. Particularly, in the paper presented by \cite{Davies16}, optical integral field data is used to separate emissions related to star formation, shock excitation, and AGN  activity in the central region of the galaxy NGC 613. The paper identifies three 'basis spectra' representing pure star formation, AGN activity, and shock excitation using BPT diagrams that can be used to distinguish the contributions in a spectrum. The study illustrates that in more than 85 percent of cases along each AGN fraction sequence, the emission line luminosities are effectively modelled using linear superposition of the luminosities from one spectrum dominated by AGN activity and another dominated by star formation processes.

As another example, \cite{cao2022} examines a broad-line `AGN' that was mistakenly classified as an H II galaxy in the BPT diagram. This misclassification is thought to be caused by contamination from SF activities. The issue is addressed by subtracting an average star formation activity from the spectrum. Their study suggests that at least 20\% of the star-forming contributions should be considered in the spectrum to rectify the misclassification, which may be a consequence of the limitations inherent in single-fibre spectroscopy.

IFU surveys have emerged as a key tool to address the limitations of single-fibre spectroscopic surveys like SDSS in capturing the complex internal structure of galaxies. IFU surveys capture data across smaller two-dimensional fields, providing a more nuanced view of the internal structures of galaxies. This approach complements the broader insights offered by surveys like SDSS. A widely utilized IFU dataset is the Mapping Nearby Galaxies at Apache Point Observatory (MaNGA) survey \citep{Bundy2015}, which provides us with less contaminated SF and NonSF regions. This makes it an exceptional training set for a range of analytical projects. Moreover, the MaNGA survey has significantly contributed to research into the spatially resolved properties of galaxies. This includes the evaluation of the NonSF fraction and its distribution across various galactic regions, a topic comprehensively explored in the study by \cite{Belfiore2019}.

In this current work, we present a decomposition method that utilizes the BPT diagram along with deep-supervised models to separate galaxy spectra into their constituent NonSF and SF components. This method is refined by training data derived from MaNGA spectra. In this paper, we describe the datasets used in this work in section \ref{sec:data}. Section \ref{sec:method} describes the method, and the results will be described in section \ref{sec:results}. An overview and the conclusion will be presented in section \ref{sec:summary}. 

\begin{figure}[ht]
    \centering
    \includegraphics[width=8.cm]{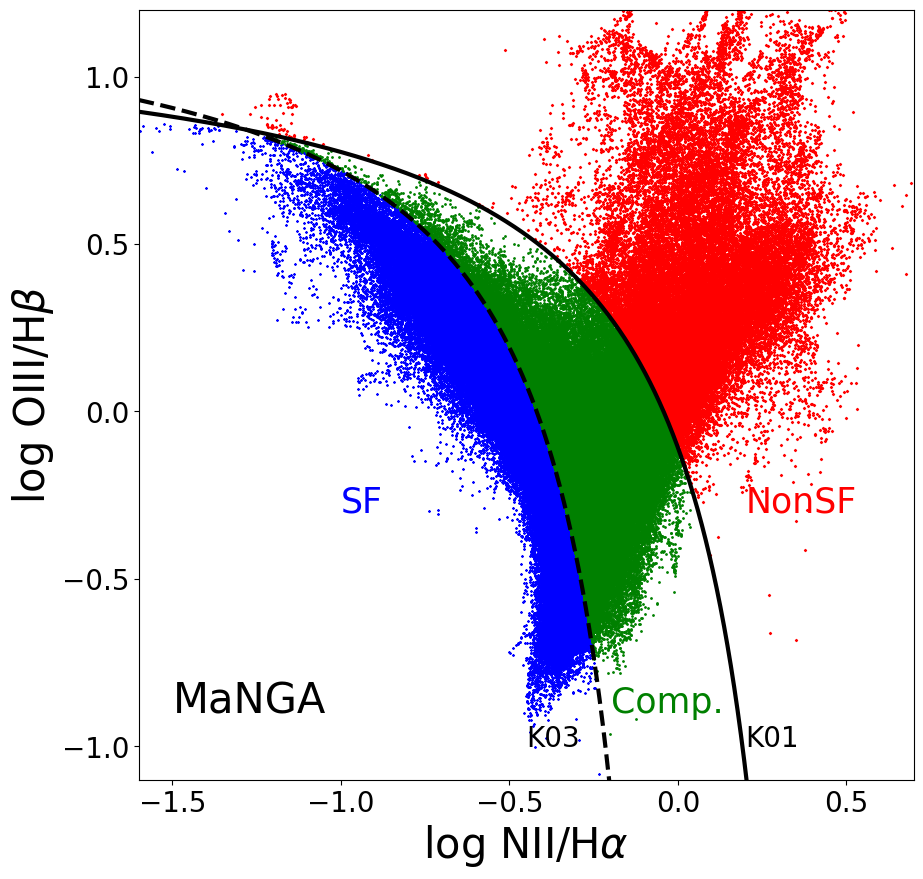}
    \includegraphics[width=8.cm]{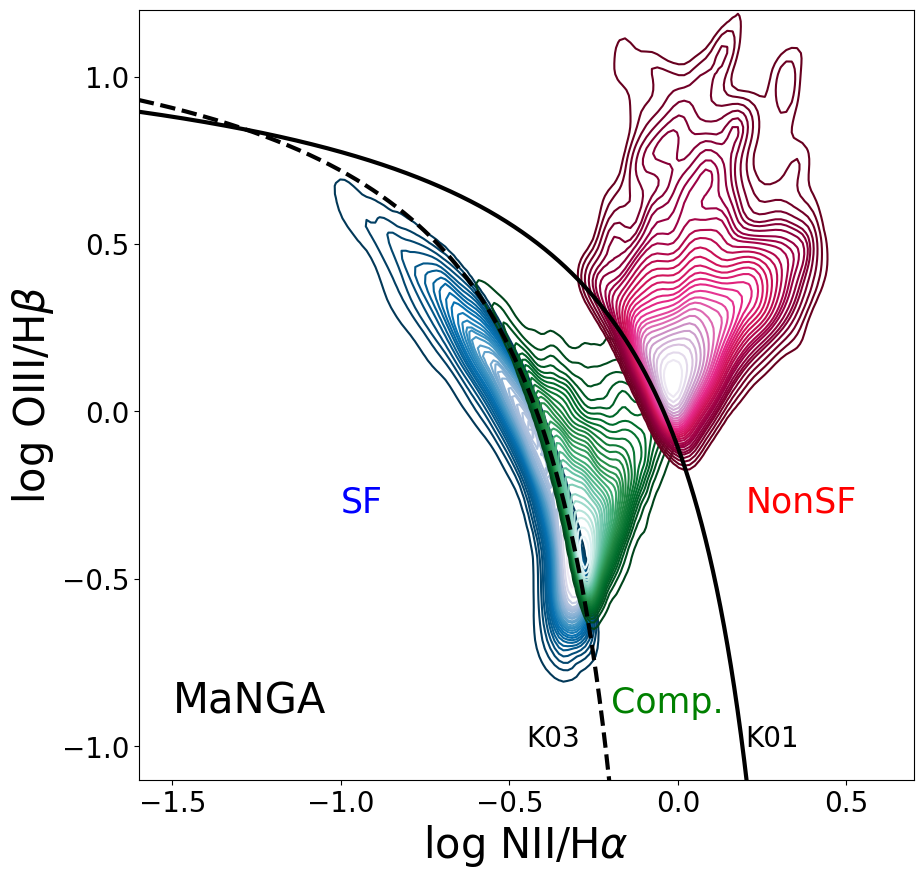}
    \caption{The top BPT diagram places MaNGA spectra on this plane based on the ratio of their emission lines. Blue points indicate typical SF spectra, red shows spectra dominated by NonSF and green shows composite spectra.
    The lower BPT diagram shows a density plot for the same MaNGA dataset to better visualize the density of spectra. Both figures only show MaNGA spectra with a $\rm{SN}>8$.}
    \label{fig:fig-BPT-manga}
\end{figure}

\section{Data}
\label{sec:data}

\subsection{MaNGA preprocessing}

Training our models for spectra decomposition involves generating synthetic spectra with known contributions from near-homogenous sources. This section outlines how we processed MaNGA spectra to act as these representative sources.

The preprocessing pipeline builds on \cite{teimoorinia2022}, and we would direct the reader there for a more thorough explanation. Briefly, the MaNGA spectra are spatial pixels (spaxels) from Data Release 15 (DR15, \cite{Bundy2015}). Spaxels with continuum signal-to-noise ratio (SN) $<$ 2 were masked, along with foreground stars.
Spaxels were spatially binned following \cite{Cappellari2003a} and \cite{Woo2019} into ``baxels'' to increase SN, accounting for correlated noise in adjacent spaxels following \cite{Law16}. Binning targeted a SN of 20 and this was not always achieved, especially for baxels towards the edge of the galaxies. The preprocessing left 4682 galaxies and 1,170,539 baxels.

For this work, baxels were resampled to 2\AA ~wavelength bins between 3700\AA ~and 8114\AA ~(therefore, each baxel is defined by 2208 elements). Due to deredshifting, some wavelength bins did not have a signal in this range, and they were removed, resulting in 1.08 million spectra.

As this paper intends to decompose emission lines into either SF or NonSF sources, we further filter baxels by signal-to-noise of their emission lines; all the major emission lines (\Ha, \Hb, \O3, \N2 and \S2) need to have SN values greater than 8. There are 1,082,389 samples before filtering, and there are 406891 samples after filtering.

We use K01 (\cite{Kewley2001}) and K03 (\cite{Kauffmann2003b}) as the lines to separate the AGN and SF samples, which is shown in Figure \ref{fig:fig-BPT-manga}. There are 41,974 NonSF samples, 173,235 SF samples, and 191,682 composite samples.  For data normalization, we divide each spectrum by its median value.

\subsection{SDSS preprocessing}
\label{sec:val_samp}

We generate a test set from integrated galaxy spectra taken from the 7th data release of SDSS (DR7). Spectra included must correspond to SDSS objects with spectroscopic classes of 2, 3, or 4 (galaxies, quasars, or high-redshift quasars, respectively) and spectroscopic redshifts below 0.3. In order to select optical/ narrow-line region AGN hosts and galaxies without a strong optical AGN contribution with high confidence, we use the MPA-JHU catalogue of spectroscopic data products for DR7 galaxies. Following our method for the MaNGA training set, we require a minimum signal-to-noise ratio of 8 for all four narrow emission lines used for BPT classification. We select samples of star-forming, composite, and NonSF host galaxies, again using the K01 and K03 diagnostics on the diagram to denote the three main regions.

The spectra for the objects in the validation set are pre-processed in the same way as the MaNGA training spectra, with additional features added to account for specific systematic issues with SDSS DR7 spectra. Inspection of the data reveals that there is a common artifact in the SDSS spectra between 5570- 5590Å produced by the over-subtraction of a telluric line. We flattened each spectrum by the local flux of the continuum in this 20Å window to prevent it from affecting the dynamic range of the data later.

As with the MaNGA data, we shift the wavelengths of each spectrum measurement to the rest frame based on the SDSS spectroscopic redshift. We used the \textsc{pysynphot} \citep{pysynphot} package to re-sample each spectrum on an equal-spaced linear grid of wavelength between 3700-8114Å. \textsc{pysynphot} conserves total flux in the spectrum, such that emission line fluxes are retained even when specific spectral features are down-sampled during data preparation.

Where data was missing on either the short- or long-wavelength side of the spectrum, a linear regression and extrapolation was performed on the spectral data below 4000Å and above 5000Å. Gaussian noise with $\sigma_{noise}=\frac{1}{2}\times\sigma_{data}$ (standard deviation equal to half the standard deviation of the real, nonzero data in the relevant wavelength domain) was also added to the linear extrapolations in order to match the characteristic variance of the spectrum. Although this choice is motivated by realism, we expect the particular characteristics of the noise will have no bearing on the performance of the model since noise is inherently unlearnable and no new information is being passed.

\subsection{Continuum estimations}

We extract the continuum from each spectrum by running pPXF \citep{Cappellari2017}, which is able to fit both the continuum and emission lines simultaneously.  We used the E-MILES templates \citep{Vazdekis2016} in a grid of 16 age bins (0.03, 0.05, 0.07, 0.09, 0.2, 0.4, 0.7, 1, 1.5, 2, 3, 5, 7, 9, 11 and 13.5 Gyr) and 10 bins of metallicity ([Z/H] = -1.79, -1.49, -1.26, -0.96, -0.66, -0.35, -0.25,  0.06,  0.15 and 0.26).  We used the option to regularize the best-fit template weights with a regularization parameter of 100 (i.e., a regularization error of 0.01).  \cite{Woo2024} found that regularization greatly improved stellar population parameter estimation for the continuum.  However, regularization is not expected to change the $\chi^2$ of the fit significantly.

\section{Method} \label{sec:method}

In this section, we will describe the methods used to estimate the necessary parameters used in this paper and also our approach to decomposing galaxy spectra.

\subsection{Flux ratio estimations}

Accurate and reliable measurement of spectral flux in astronomy must account for various factors. This may involve correcting spectra for dust and other intrinsic and environmental influences \citep[e.g.,][]{Calzetti-2001}. Additionally, when measuring physical parameters, the impact of utilizing different methods and templates \citep[e.g.,][]{cao2022}, along with the significant role of spectroscopic techniques and data processing algorithms \citep[e.g.,][]{Bolton-2012}, are crucial. These factors can lead to variations between different studies or surveys. Our study utilizes spectra from the SDSS and MaNGa surveys, which already have corresponding flux measurements. In a machine learning context, these data sets can potentially be used to train deep neural networks to obtain fluxes. Since the spectra are generated synthetically by a neural network, the flux values are no longer known. In the next phase of our work, we will need to handle such spectra generated by our pipeline, which will require new flux measurements. This presents a situation where we need to measure fluxes from different sources. So, we need a method for obtaining the measurable parameters from different sources required to construct BPT diagrams in a consistent and uniform way.

\begin{figure}[ht]
    \centering
    \includegraphics[width=8.cm]{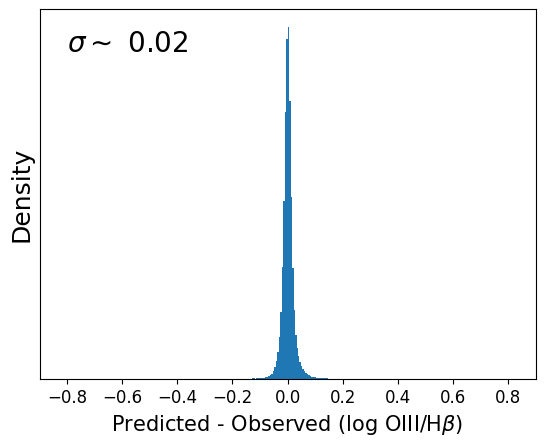}

    \caption{The difference between predicted log(OIII/H$\beta$) and the observed one.}
    \label{fig:fig-Ratio-predicted}
\end{figure}

One potential machine learning solution for measuring flux could involve training a deep neural network (DNN) with a specific dataset, such as SDSS. For example, the trained model can be used to estimate MaNGA spectra fluxes. Initially, after training a model, this method produces good results when validated against the SDSS validation set. However, when we examine the results, it exhibit bias when validated with MaNGA spectra.  A bias in SDSS flux estimations is also observed when MaNGA spectra are used as the training set. Even when a combined training set of SDSS and MaNGA is employed, individual validation of each survey shows biases in the validation process, with one showing bias to the left and the other to the right. This might indicate the different nature of the datasets or data processing procedures or the effect of environmental factors such as dust. In the next step, we utilize the ratio of flux measurements as the target instead of directly measuring the flux. This eliminates systematic issues and prevents biases from being present. In practice, since we need the ratio of the fluxes, we train and use a DNN to measure flux ratios directly (hereafter, ratio-DNN). The comparison between the observed and predicted ratios shows very little scatter, and no bias is seen in different validation processes; therefore, the classification of spectra by BPT diagrams (SF, NonSF, and composites) achieves very high accuracy. Figure \ref{fig:fig-Ratio-predicted} depicts a validation set showing the difference between the predicted log(OIII/H$\beta$) and the observed value. For consistency and uniformity, we utilize this ratio-DNN to make our BPT diagrams.

The ratio-DNN model is a regressor, which means the output values are real numbers instead of labels used for classification tasks. This model inputs 2208 data points from a spectrum. It includes eight one-dimensional Convolutional Neural Network \citep{Gu2015} layers. These layers can extract useful features and information from the input spectra. The last layer of the CNN part is connected to a fully connected regressor \citep[e.g., ][]{Kohler19}, ending with nodes related to flux rations needed to construct BPT diagrams. To avoid overfitting, we employ various regularization methods such as Dropouts \citep[e.g.,][]{Liang2021} and an Early Stopping method that halts training if the validation set performance drops below that of the training set.

\subsection{Spectra Continuum Estimations}

In this paper, we investigate two distinct scenarios. Our approach to analyzing spectra yields results with both the original spectra and the same spectra from which the continuum has been subtracted. One can train various DNNs (as a pipeline) using the original spectra and adopt a more data-driven approach that minimizes dependence on theoretical assumptions and models. However, since NonSF emission lines typically lie in regions with very different stellar populations than SF emission lines, one potential concern is that the DNN will learn from the continuum rather than the emission lines.  Therefore, we also repeat the analysis by training another set of DNNs with continuum-subtracted spectra. Our research indicates that using either the original or the continuum-subtracted spectra produces similar results. 

\begin{figure*}[hbt!]
    \centering
    \includegraphics[width=15.cm]{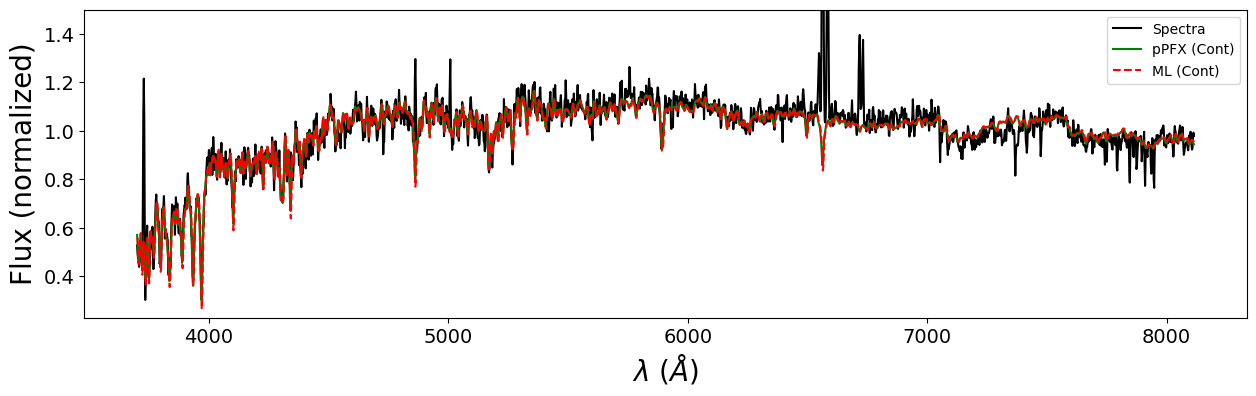}
    \includegraphics[width=15.cm]{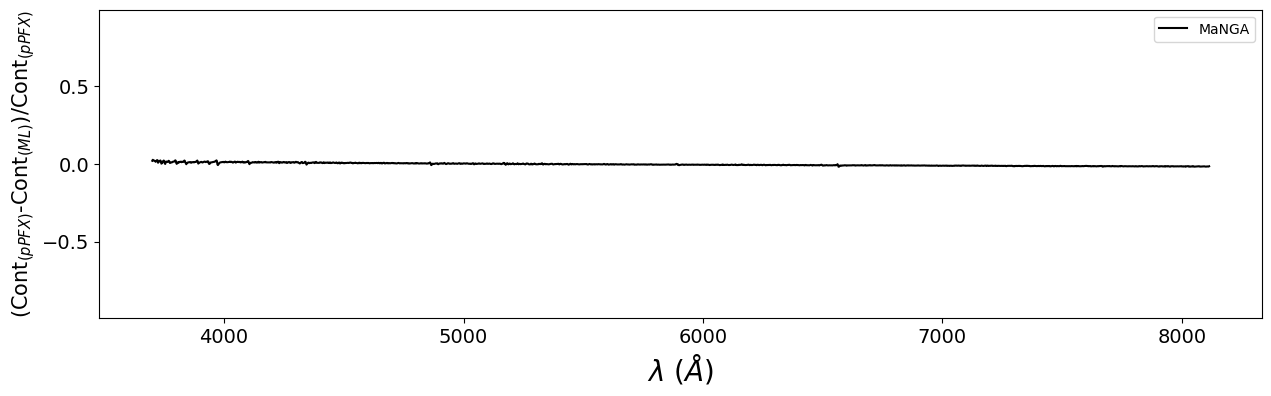}
    \includegraphics[width=15.cm]{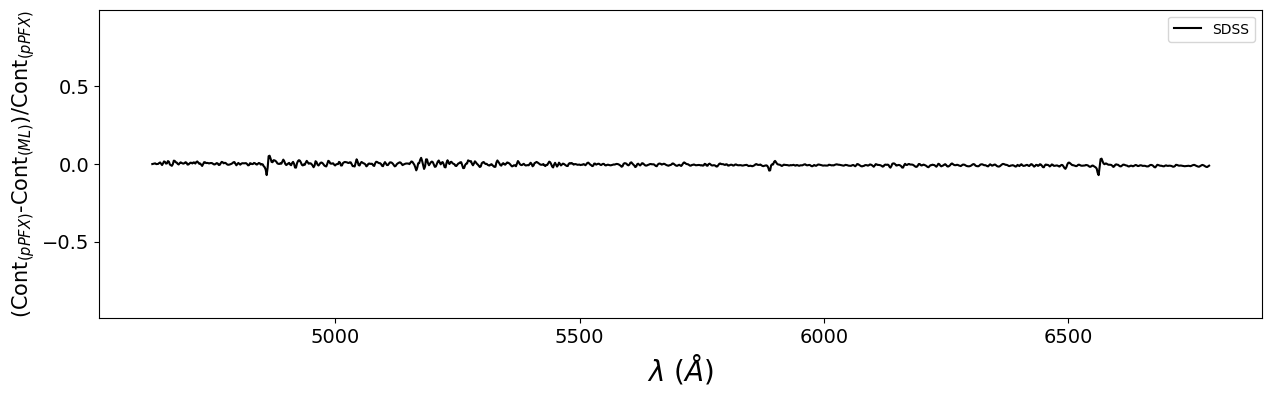}
    \caption{A random spectrum is depicted in black on the top plot, which also compares the continuum estimated by pPFX (shown as a green solid line) to that predicted by ML (shown as a red dashed line). The middle plot illustrates the average difference between the pPFX and ML estimates for MaNGA spectra across 50,000 samples. The bottom plot displays the same comparison for SDSS spectra.}
    \label{fig:fig-continuum}
\end{figure*}

Similar to the previous section, our method requires us to work with spectra generated by DNNs and estimate the continuum when necessary. To achieve this, we use a machine learning model, a DNN (hereafter, cont-DNN), where the input is the spectra, and the predicted outputs are the best-fitted continuum. It should be noted that the target for the deep model is the continuum estimated by pPFX; thus, the ML model mimics pPFX in estimating the continuum. Using the data and continuum described in Sec. \ref{sec:data}, we train our cont-DNN to obtain a continuum for each input spectrum. The results are unsurprisingly excellent, largely because training a deep model on noiseless theoretical models tends to converge quickly and produces outputs with high accuracy, as demonstrated here. The top plot of Figure \ref{fig:fig-continuum} shows a spectrum alongside the continuum by pPFX and the continuum generated by cont-DNN. The middle plot illustrates the relative average difference between the pPFX and ML estimates for MaNGA spectra from 50,000 samples. The bottom plot illustrates the same comparison for SDSS spectra. The results are nearly perfect for MaNGA spectra, with slightly more fluctuation for SDSS spectra.  Note that although the MaNGA and SDSS datasets were constrained to have the same limiting S/N for the emission lines, they were not constrained to have the same S/N for the continuum.  MaNGA baxels were binned to achieve a continuum S/N of 20, while there was no such binning or constraint on the continuum S/N for the SDSS.

\subsection{Artificial spectra} 
\label{sec:ArtificialComposites}

Our method for decomposing an unknown spectrum into SF and NonSF components involves synthesizing a set of known artificial samples. A DNN is then trained to reverse this process. MaNGA spectra have been classified into three regions based on their locations on a BPT diagram (see Figure \ref{fig:fig-BPT-manga}). A uniformly random value between 0 and 1 is selected as the weight for a randomly chosen SF spectrum, with the remaining fraction applied to a randomly selected NonSF spectrum. These two 'source' spectra are weighted and summed to create a synthetic spectrum. We record both source spectra and their weights as the training dataset. To create the training set, we only utilize 'pure' SF and 'pure' NonSF spectra. Below, we provide two examples of the procedure.

\begin{figure*}[hbt!]
    \centering
    \includegraphics[width=18.cm]{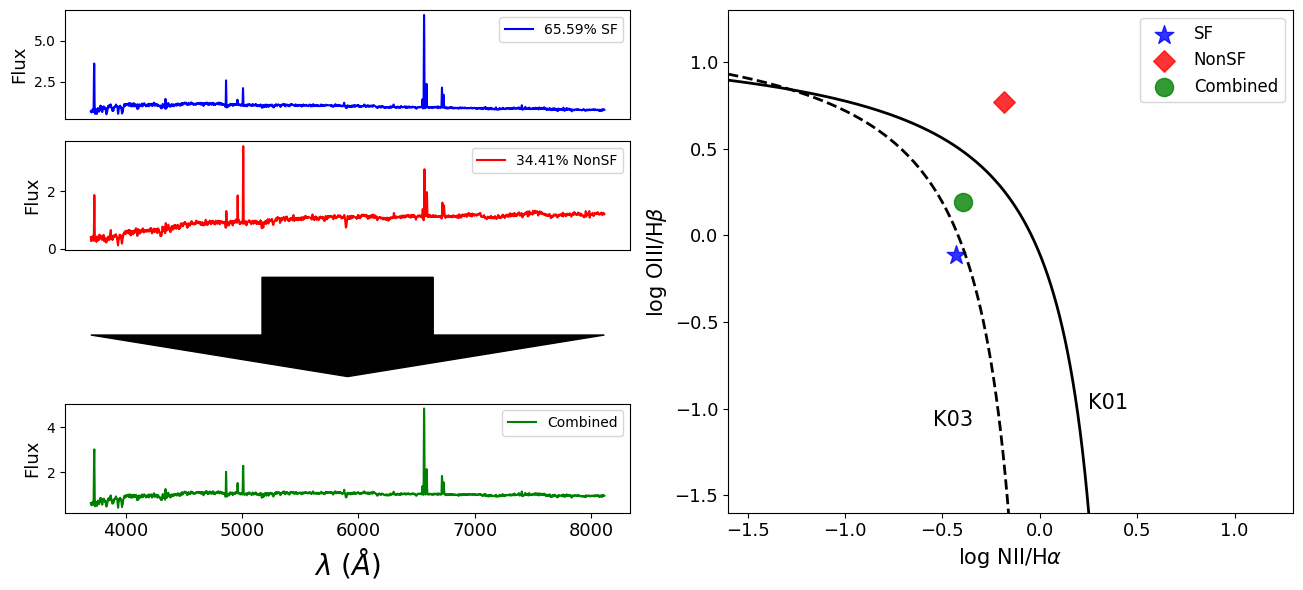}
    \caption{The left panel displays two arbitrarily chosen spectra: one from the (MaNGA) SF region (depicted in blue) with an assigned weight of $0.6559$, and the other from the NonSF domain (depicted in red) with a weight of $0.3441$. The synthetic spectrum is then calculated as the weighted average of these NonSF and SF spectra, as illustrated in the bottom left. The right panel shows the positions of the three spectra on the BPT diagram.}
    \label{fig:fig-combine-BPT-01}
\end{figure*}

The left panel of Figure \ref{fig:fig-combine-BPT-01} provides a visual demonstration, with an SF spectrum (depicted in blue) assigned a weight of $0.6559$ and a NonSF spectrum (in red) assigned a weight of $0.3441$. The resultant combined spectrum is calculated as the weighted average of these individual NonSF and SF spectra, as shown at the bottom left of the figure. The right panel displays the positions of the two source spectra on the BPT diagram and the placement of the resultant synthetic spectrum. As can be seen, this synthesized spectrum is situated between the K01 and K03 demarcations, a positioning influenced by the randomly selected SF and NonSF sources and their respective weights. Different weights or initial SF/NonSF spectra will result in synthetic spectra that can occupy various regions on the BPT diagram.

\begin{figure*}[hbt!]
    \centering
    \includegraphics[width=18.cm]{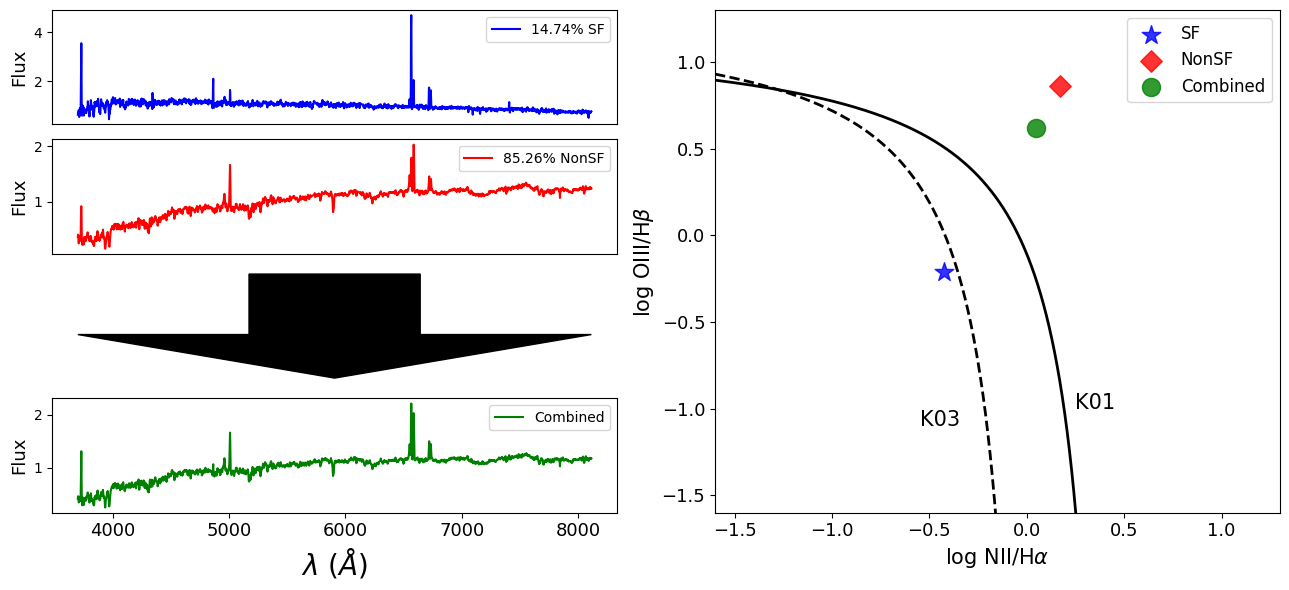}
    \caption{
This is similar to Figure \ref{fig:fig-combine-BPT-01}, where two randomly selected and weighted spectra are combined to form a synthetic composite. Due to the location and weighting of these initial spectra, the resulting spectrum is positioned within the NonSF region.}
    \label{fig:fig-artificial-comp-BPT-02}
\end{figure*}

\begin{figure}[hbt!]
    \centering
    \includegraphics[width=8.cm]{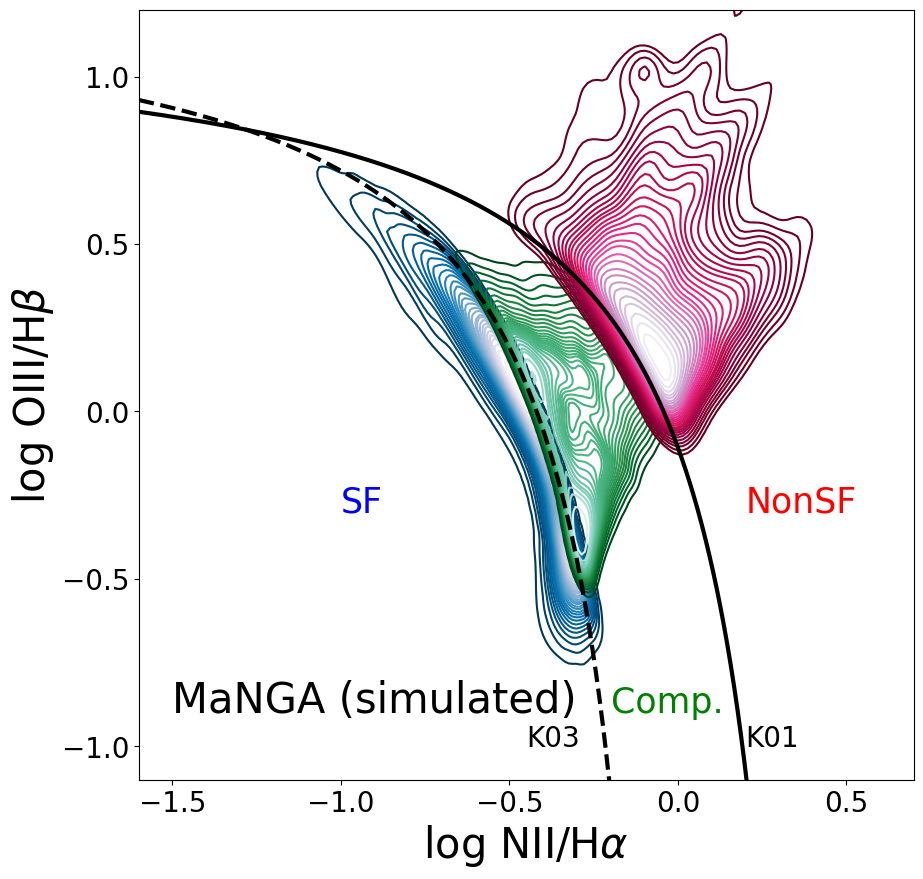}
    \caption{A subset of 50,000 synthetic spectra is displayed here to illustrate their distribution. Similar to Figure \ref{fig:fig-BPT-manga}, these spectra can be classified into the three regions of the BPT diagram and coloured accordingly.}
    \label{fig:fig-FluxAll-manga}
\end{figure}

In Figure, \ref{fig:fig-artificial-comp-BPT-02}, we present a scenario where the synthetic spectrum is positioned not between the division lines K01 and K03 but within the NonSF region. This placement demonstrates that the location of the combined spectrum—whether between K01 and K03 or outside this range—depends on the specific weightings and source spectra used. Our training set includes one million synthetic spectra, supplemented by an independent validation set of half a million spectra. A subset of 50,000 synthetic spectra is displayed in Figure \ref{fig:fig-FluxAll-manga} to illustrate their distribution. Similar to Figure \ref{fig:fig-BPT-manga}, these spectra can be classified into the three regions of the BPT diagram and coloured accordingly. We record the weights and initial spectra used to create each synthetic spectrum to train our decomposition model.

\subsection{The Deep Decomposition Model}
\label{sec:DecompositionSystem}

The main purpose of generating synthetic spectra is to use them as known targets in training a decomposition model. Essentially, the process executed by our Deep Decomposition Model (hereafter, DDM) reverses the method used to create the synthetic spectra. Our model comprises a neural network with five distinct outputs, each governed by its own loss function. Detailed information on the structure of all the models and the pipeline can be found \href{https://github.com/shishehchi/SpectraDecomposition}{\underline{here}}. 

From a technical perspective, the model comprises eight consecutive one-dimensional CNN layers to extract valuable features and information from the input spectra. This is followed by five distinct individual regressors corresponding to five outputs. The overall trainable parameters are about 39 million, and the model is evaluated with half a million independent synthetic spectra as validation sets and employs various regularization methods such as Dropouts and the Early Stopping method.

In summary, the DDM processes a combined spectrum featuring 2,208 wavelength bins and generates five distinct outputs. Among these, two are spectra: one representing the SF component and the other the NonSF component, each consisting of 2,208 data points. In addition to these spectra, the DDM produces three additional outputs. One of the outputs predicts the SF contribution as a fraction, indicating the proportion of the SF component within the combined spectrum (or 1-SF for NonSF). The other two outputs are the flux ratios, expressed in a logarithmic scale, that are needed to create the BPT diagram. These outputs (the ratios) help the model learn and focus on predicting the proper positions needed to construct the BPT diagram.

To maintain consistency in terminology, we use the term 'composite' spectra to refer to those that fall between the K01 and K03 lines on a BPT diagram, while 'combined' spectra refer to the input to our model. A combined spectrum may also be a composite.

\begin{figure*}[hbt!]
    \centering
    \includegraphics[width=18.cm]{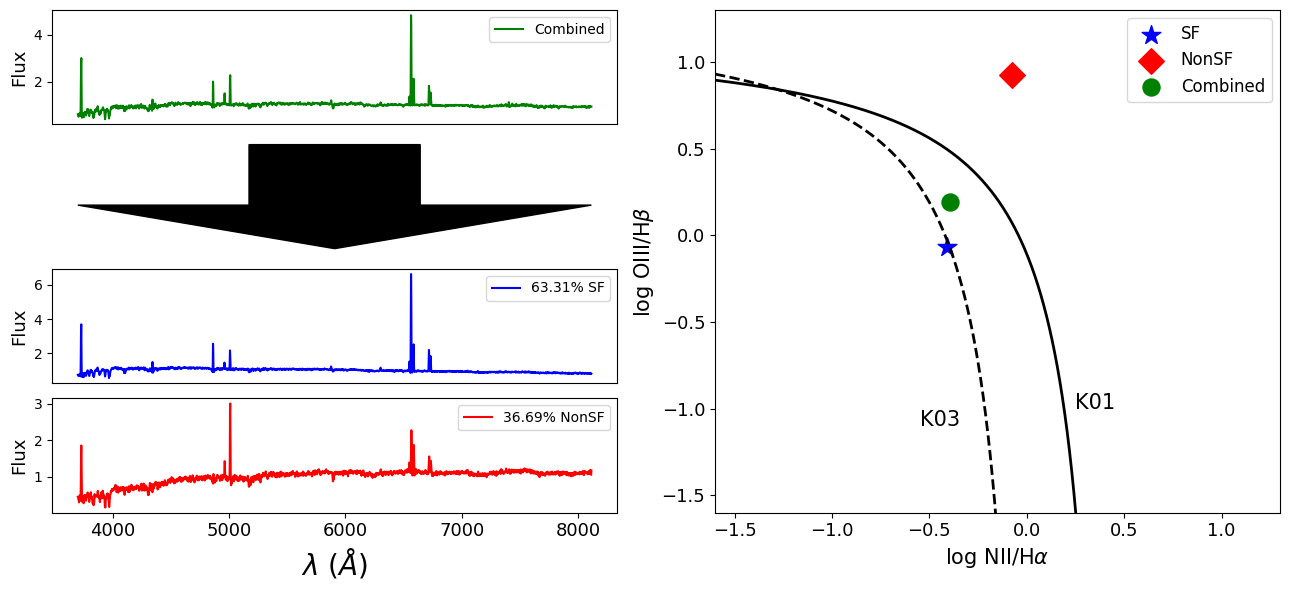}
    \caption{This plot illustrates an example of a combined spectrum, shown in green, being fed into the model, which then generates two spectra: SF and NonSF, each with corresponding contributions. This example serves as the inverse of Figure \ref{fig:fig-combine-BPT-01}.}
    \label{fig:model_output-1}
\end{figure*}

An example model output is displayed in Figure \ref{fig:model_output-1}. The left panel features the input combined spectrum (shown in green) being processed by the model. The predicted SF (in blue) and NonSF (in red) spectra, along with their corresponding contributions, are also illustrated. The right panel displays the placement of the two predicted decomposed source spectra on a BPT diagram alongside the original combined spectrum. For comparison, this example serves as the inverse of Figure \ref{fig:fig-combine-BPT-01}. As can be seen, the SF contribution to the predictions is in good agreement with expectations.  Here, we see that the SF component spectrum exhibits higher $\rm{H}\alpha$ intensity in comparison to the combined spectrum. This might raise concerns about flux conservation and, consequently, the conservation of the star-formation rate. However, when we consider the contribution percentage of each spectrum, the total flux remains conserved.

\begin{figure}[hbt!]
    \centering
    \includegraphics[width=9.cm]{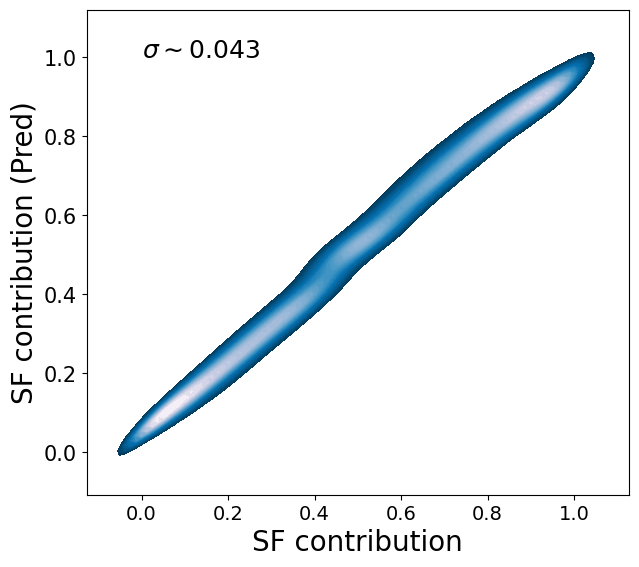}
    \caption{This plot shows the total scatter between the SF contribution in making the synthetic spectra and the predicted ones. A sample of 50,000 predictions is compared with the true values.}
    \label{fig:fig-sf}
\end{figure}

The total scatter between the SF contribution in making the synthetic spectra and the predicted ones is $\sigma \sim 0.043$. A sample of 50,000 predictions is compared with the true values in Figure \ref{fig:fig-sf}.

\section{Results}
\label{sec:results}

\subsection{Flux Conservation}

\begin{figure*}[hbt!]
    \centering
    \includegraphics[width=16.cm]{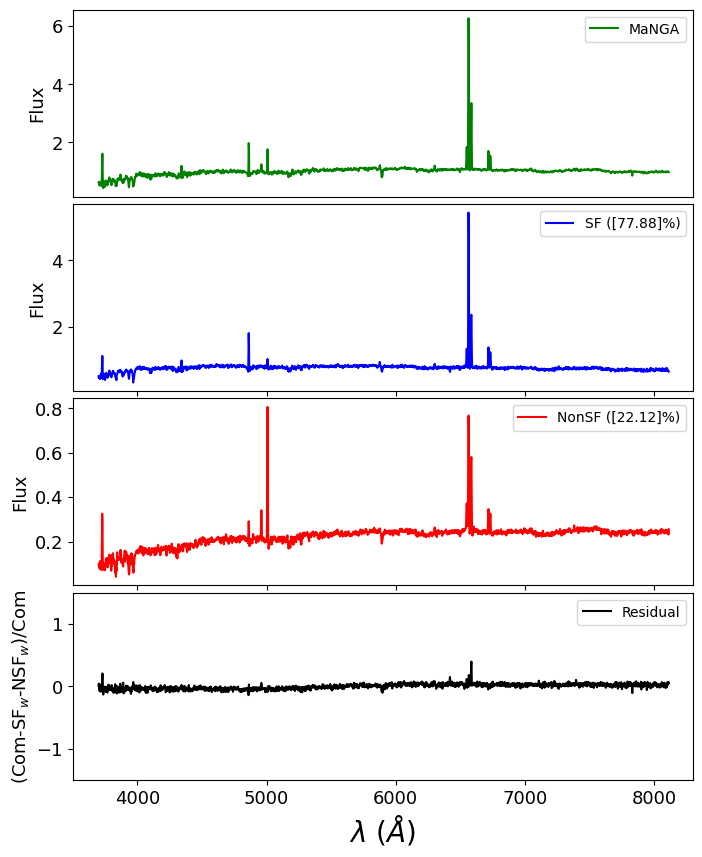}
    \caption{The input to our model is displayed in the top panel. The middle panels present the two output components along with their corresponding weights. The lower panel illustrates the relative difference between the input and the sum of the weighted outputs.}
    \label{fig:fig-Diff-Flux-Simulated}
\end{figure*}

Flux conservation is an important factor in assessing the quality of our model.  With robust flux conservation, the decomposed SF spectrum can be used to extract emission line fluxes and calculate important physical parameters like SFR. In Figure \ref{fig:fig-Diff-Flux-Simulated}, we present a single example of flux conservation where an input spectrum (from the synthetic test set) shown in the top panel is decomposed into two spectra displayed in the middle panels. These panels also indicate the fraction contributed by each spectrum. By taking the weighted average of the two predicted sources, we can reconstruct the input. The relative residuals between the reconstruction and the original input are shown in the bottom panel, where near-zero variance along the wavelength bins indicates that flux conservation is only minimally perturbed at emission lines. While this is just one example, we will explore creative average flux conservation in the following.

\begin{figure*}[hbt!]
    \centering
    \includegraphics[width=8.cm]{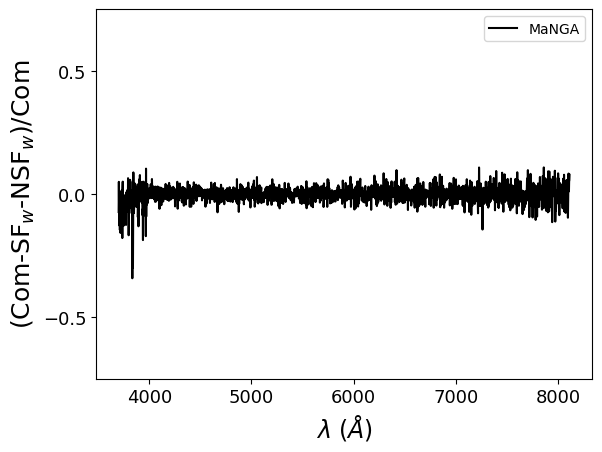}
    \includegraphics[width=8.cm]{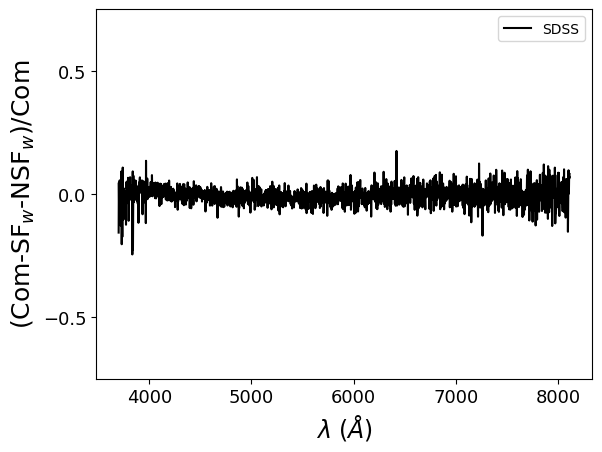}
    \includegraphics[width=8.cm]{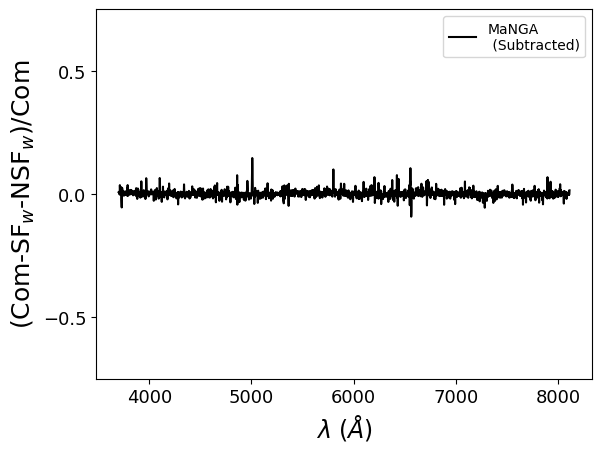}
    \includegraphics[width=8.cm]{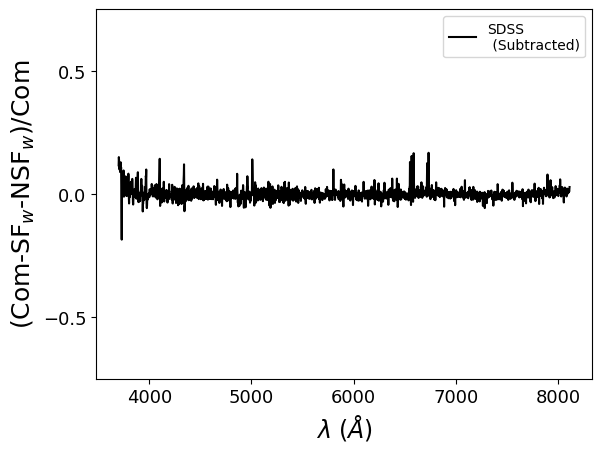}
    \caption{The relative residuals for the MaNGA and SDSS datasets (for a sample of 50,000 'original' spectra) are shown in the top panel, while the lower panel displays the same for continuum-subtracted spectra that are used to train the associated DDM.
    }
    \label{fig:residual-combined}
\end{figure*}

Figure \ref{fig:residual-combined} displays the relative average residual for 50,000 decomposed spectra, separated by wavelength bin. The reconstructed spectrum is the sum of the predicted SF/NonSF spectra, weighted by their predicted contributing fractions. The upper panel presents the average residuals for MaNGA and SDSS spectra for the original spectra (i.e., not continuum-subtracted). It shows a fairly flat residual with negligible fluctuations at the emission line positions. MaNGA spectra show smaller residuals due to MaNGA baxels being binned to achieve S/N for the continuum of at least 20, while SDSS spectra were not binned.  However, both cases exhibit a bit more fluctuation at wavelengths less than 4000 $\AA$. On average, this panel indicates good flux conservation, especially at wavelengths greater than 4000 \AA, which is crucial for constructing BPT diagrams. 

The lower panel pertains to the continuum-subtracted spectra, which are used to train the associated DDM. These show less residual at wavelengths below 4000 Å, likely due to the flat spectra (continuum-subtracted) used in training the associated DDM. However, they exhibit higher fluctuations at the positions of emission lines, which is more noticeable for the SDSS dataset.  The lower average continuum S/N in the SDSS dataset likely resulted in noisier continuum subtraction.  It should also be noted that the SDSS spectra result from the combinations of SF and non-SF emission on top of a generally bulge-dominated central spectrum, whereas the machine learning models have been trained using MaNGA spectra, which are mainly from HII regions in young disk areas and from non-SF regions in old central stellar regions. The difference could cause additional fluctuations or inaccuracies in the SDSS spectra.

\subsection{Decomposing SDSS\_J1042\_0018}

SDSS J1042-0018 is a typical broad-line AGN; however, the flux ratios of its narrow emission lines categorize it as an H II galaxy in the BPT diagram, prompting further investigation into its unique properties as discussed in various studies \cite[e.g., see][and references therein]{cao2022}. In the analysis presented by \cite{cao2022}, the emission lines of SDSS J1042-0018 are examined using two distinct models: one employs broad Gaussian functions, and the other utilizes broad Lorentz functions for the broad Balmer lines. These differing approaches lead to variant flux ratios of the narrow emission lines, resulting in the classification of SDSS J1042-0018 as an H II galaxy in the BPT diagram, even though it is actually a broad-line AGN. To explain this misclassification, one plausible theory proposes the presence of significant star-forming activity, suggesting that at least a 20\% contribution from star formation is necessary to justify the misclassification.

\begin{figure*}[hbt!]
    \centering
    \includegraphics[width=16.cm]{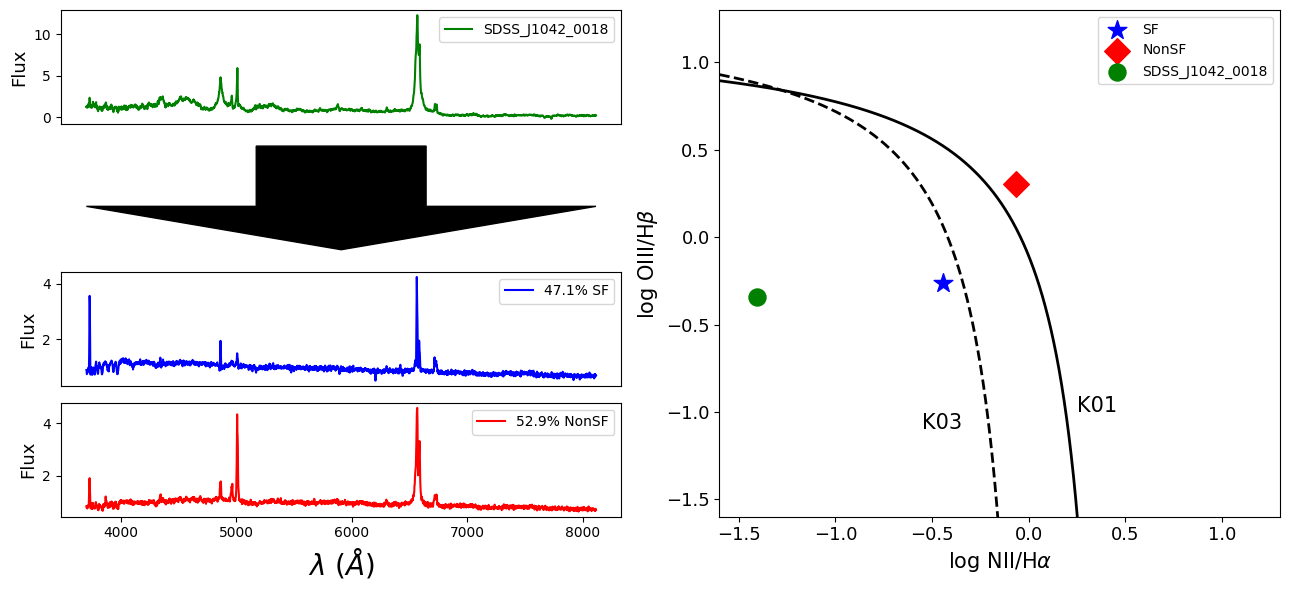}
     \includegraphics[width=16.cm]{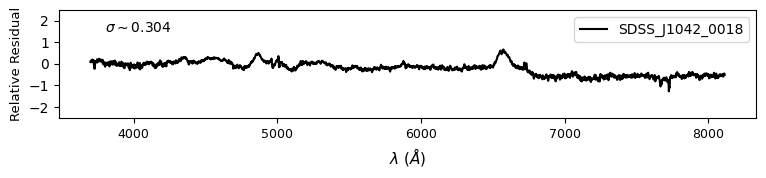}

    \includegraphics[width=16.cm]{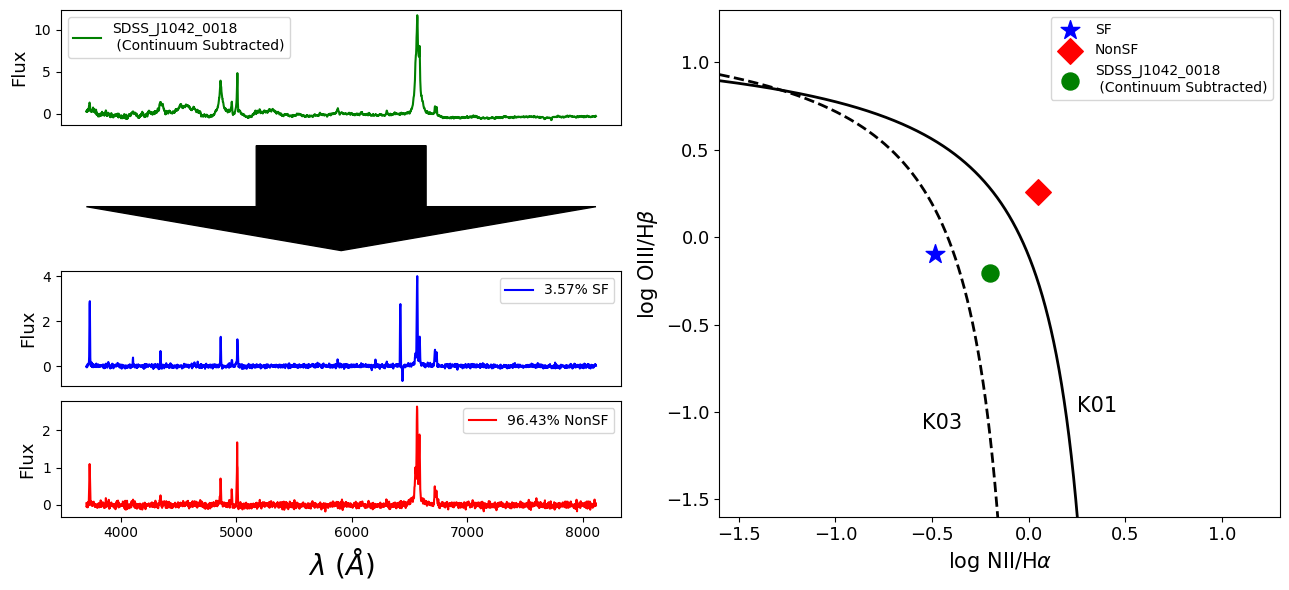}
     \includegraphics[width=16.cm]{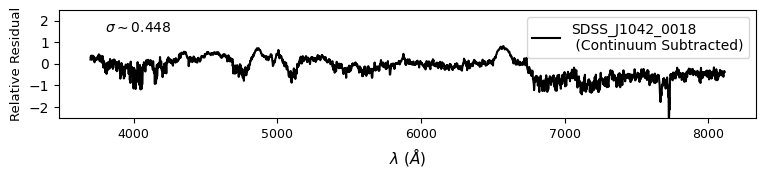}
    \caption{the plot illustrates the spectral analysis of SDSS J1042-0018 using two different deep decomposition models (DDM). The top panel displays the spectrum categorized as an 'outlier' star-forming (SF) galaxy with a significant SF contribution of 47.1\%. Noticeable fluctuations are observed in the relative residual, particularly around 4800 and 6600 Å. The third panel presents results from the model with continuum subtraction, classifying the spectrum as a composite with only a 3.5\% SF contribution. The bottom panel shows the relative residual for this scenario, marked by increased fluctuations}
    \label{fig:Decompose-J1042}
\end{figure*}

The spectrum of SDSS J1042-0018 has unique features, and visually, it is distinct from our training set and does not resemble average SDSS spectra. As a case study, we fed the spectrum into two DDM models, one trained with and the other without continuum subtraction. In the top panel of Figure \ref{fig:Decompose-J1042}, the spectrum is identified as an 'outlier' SF galaxy. The decomposed SF part contributes approximately 0.471. The associated relative residual is fairly flat; however, it exhibits high fluctuations, particularly around 4800 and 6600$\AA$. The third panel from the top shows results for the continuum-subtracted scenario. Here, the associated model classifies the input spectrum as a composite spectrum with only a 3.5\% SF contribution. The bottom panel displays the relative residual, which shows more fluctuation. In particular, the original spectra scenario (the two top panels) aligns more closely with expectations, as it shows more than a 20\% contribution from the SF part \citep[as predicted by ][]{cao2022} and also presents a better residual profile.

As described in \cite{cao2022}, parameter estimations are challenging in this case and often yield results that are not definitive. Although our results are not significantly aberrant, conclusive outcomes may not be expected due to the unique nature of this spectrum. Nevertheless, our method can still be effectively utilized for data mining and deriving insights based on a specific training set.

\subsection{Statistical Decomposition Results}

\begin{figure*}[hbt!]
    \centering
    \includegraphics[width=5.6cm]{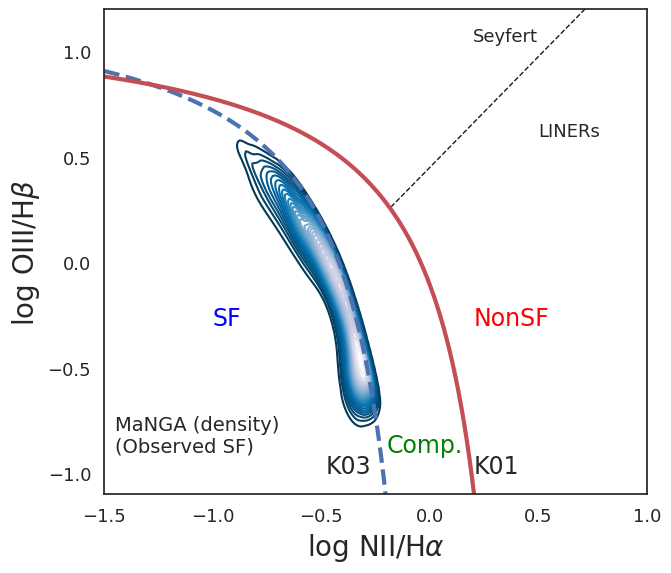}
    \includegraphics[width=6.6cm]{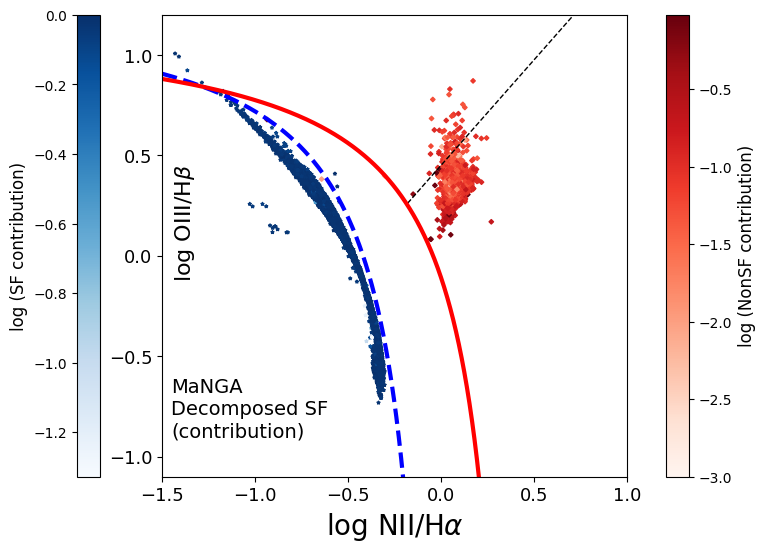}
    \includegraphics[width=5.5cm]{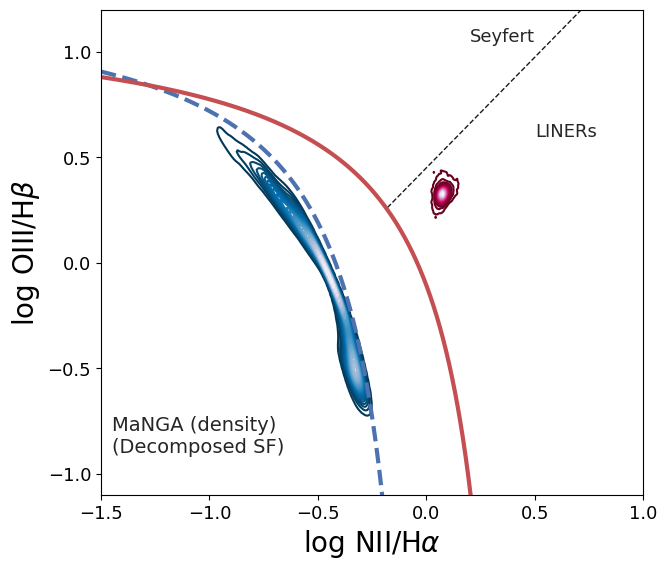}
    
    \includegraphics[width=5.6cm]{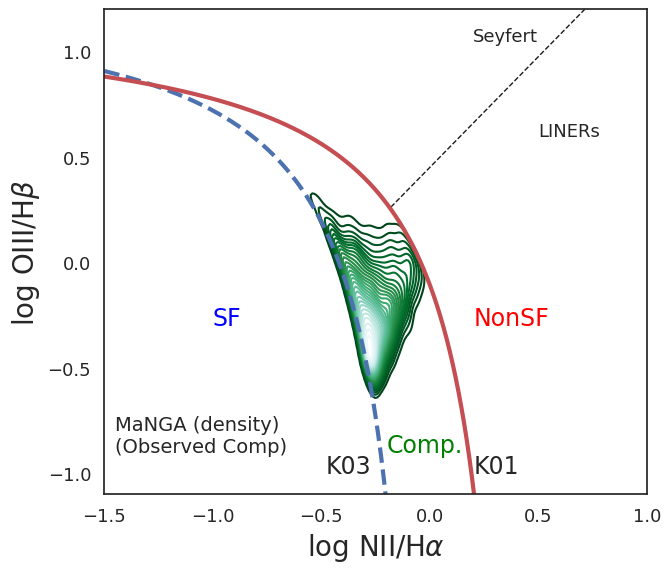}
    \includegraphics[width=6.6cm]{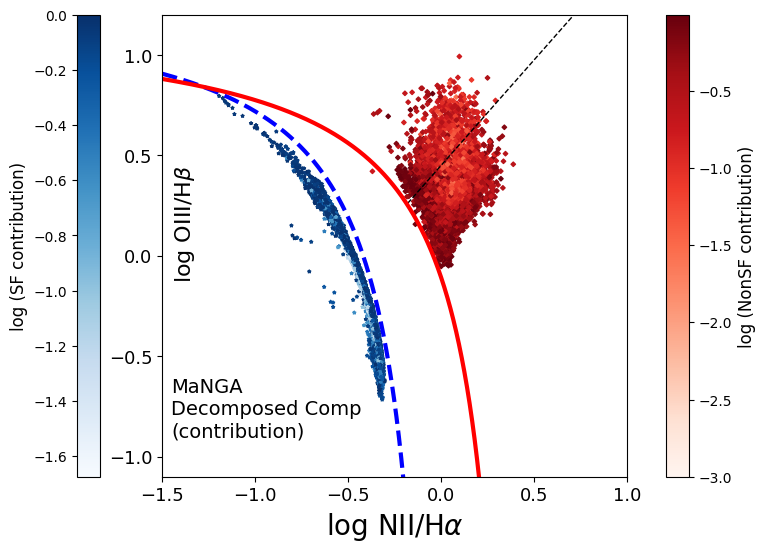}
    \includegraphics[width=5.5cm]{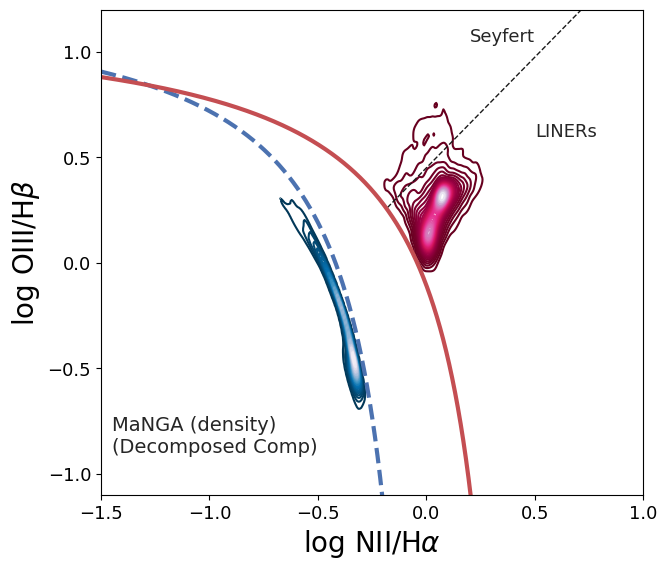}

    \includegraphics[width=5.6cm]{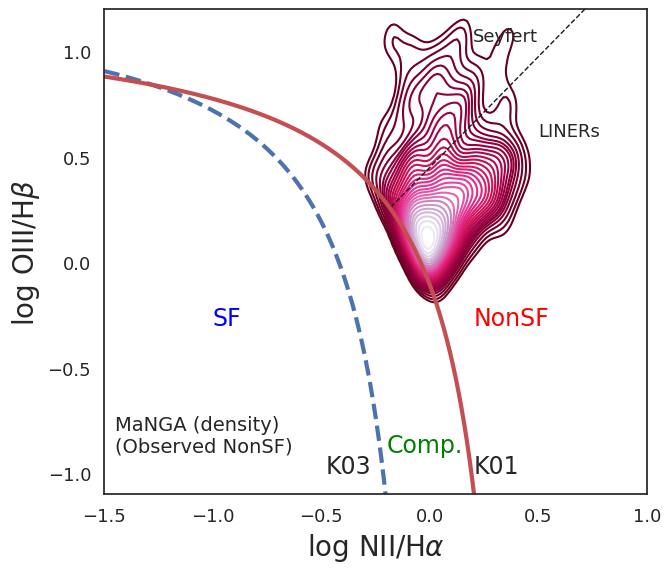}
    \includegraphics[width=6.6cm]{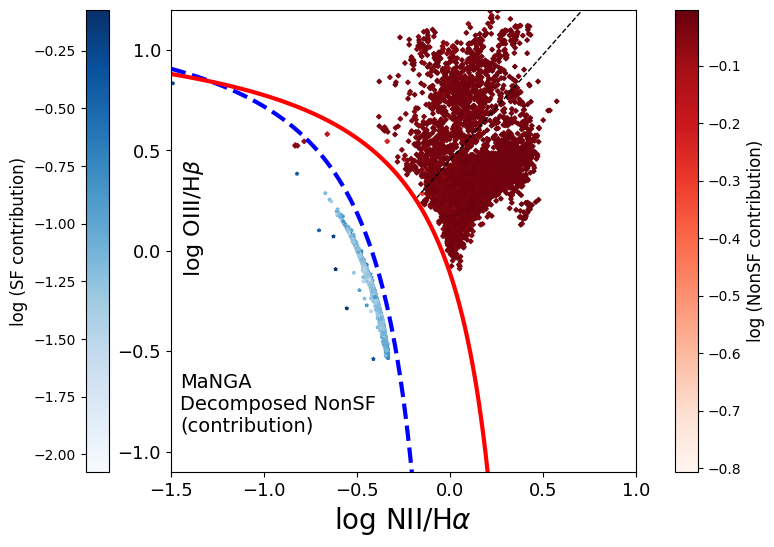}
    \includegraphics[width=5.5cm]{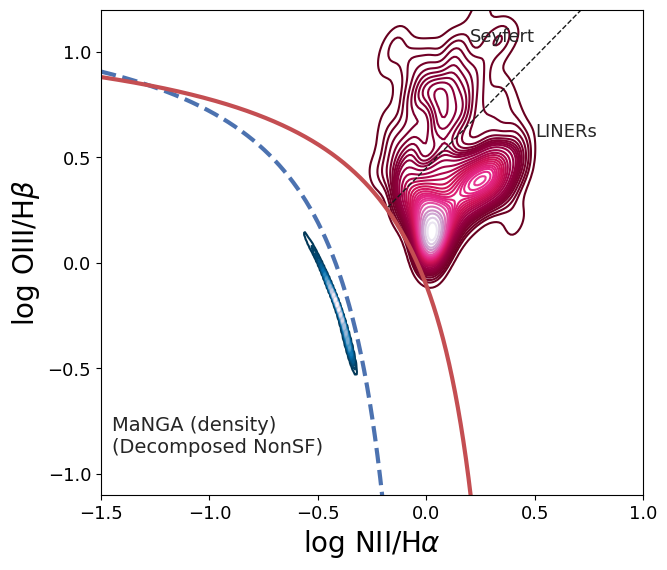}

    \caption{ On the upper panel, the top-left plot shows MaNGA SF spectra (fed to the model), revealing some NonSF contributions, which are quantified in the top-middle plot on a logarithmic scale (the top-right is the density plot). The middle panel illustrates the results of applying the decomposition model to MaNGA 'composite' spectra, highlighting extended contributions in the SF and NonSF regions. The bottom panel shows the results of the purification process applied to NonSF spectra, identifying clusters in the regions of Seyfert and LINER spectra \citep[the dotted line in the NonSF region;][]{Kewley06, Cid10}.}
    \label{fig:fig-decompose-MaNGA}
\end{figure*}

In this section, we analyze the decomposition of two datasets (MaNGA and SDSS) using large samples from different classes (SF, Composite, and NonSF). It's important to note that each spectrum, represented as a point on the BPT diagram, is split into two distinct points on a (new) BPT diagram. So, for example, if we have a sample of 50,000 SF spectra from the MaNGA dataset (shown in the top left plot of Figure \ref{fig:fig-decompose-MaNGA}), after feeding them to the model, we will have 50,000 points of NonSF and 50,000 SF on a new BPT diagram, each with different contribution values. This is depicted in the top middle plot, where the contribution values are shown on a logarithmic scale. It's evident that the Non-SF contributions are very low and mainly in the LINER sub-region. This low-value NonSF contribution is expected from IFU MaNGA spectra, which are more uniform. In other words, although we initially assumed a 'pure' SF, the results show the presence of low NonSF contributions. The top right plot displays the same result as a density plot, which shows the NonSF components concentrated in the LINER sub-region. This 'purification' process can also be extended to other classes on the BPT diagram.

In the middle panel of Figure \ref{fig:fig-decompose-MaNGA}, we feed the MaNGA `composite' spectra to the model (the left plot). These spectra are already known as a more heterogeneous set in which a more scattered distribution of  SF or NonSF is expected to be seen in the decomposition procedure.  As shown in the middle plot of this panel, SF and NonSF spectra show more extended populations and higher contributions (compared to the SF sample). In the bottom panel of the figure, the 'purification' is presented for a sample of NonSF spectra. Here, we feed a set of 'pure' NonSF spectra to the model (i.e., the bottom left plot). As expected and shown on the bottom middle plot, there is a low probability of data points appearing in the SF region. On the bottom right plots, we show the associated density plot.  This plot demonstrates that with the purification process, two more distinguishable clusters of spectra in the regions of Seyfert and LINERs are now noticeable. In this case, the density is concentrated in the LINER region potentially true Seyfert AGN emission is quite rare in nearby galaxies and may not be present in the MaNGA sample. As we will observe, the clustering pattern becomes quite clear when we test the models using the SDSS dataset, with a dense population in the Seyfert region.

\begin{figure*}[hbt!]
    \centering
    \includegraphics[width=5.8cm]{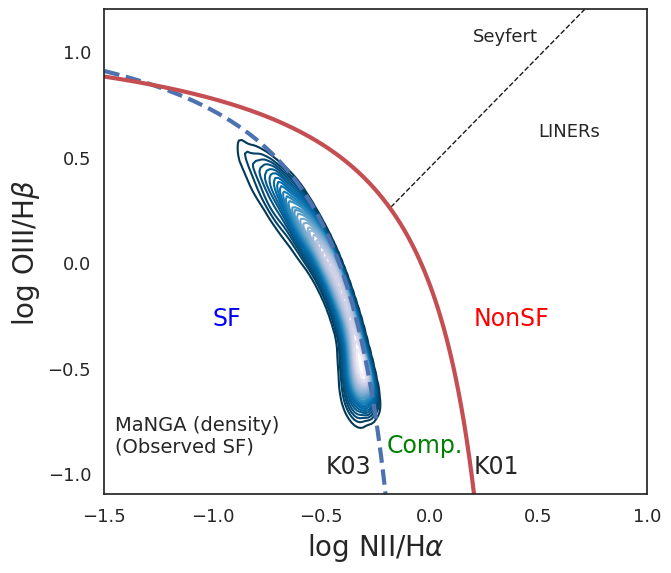}
    \includegraphics[width=6.6cm]{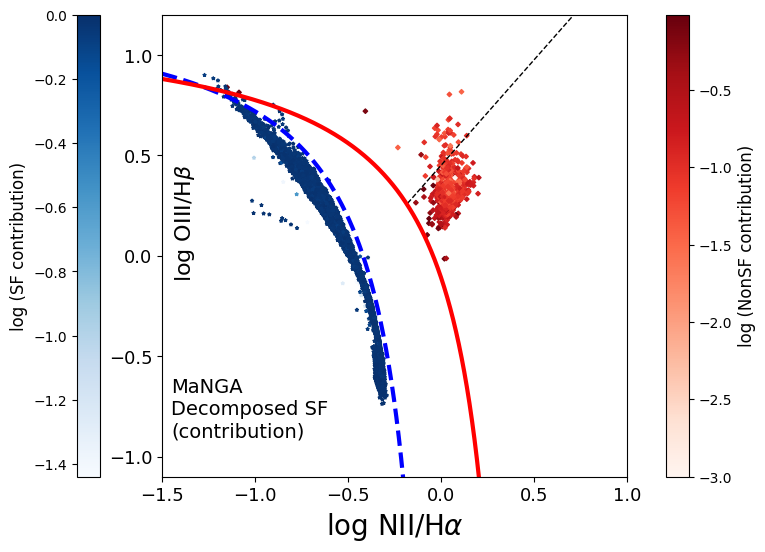}
    \includegraphics[width=5.4cm]{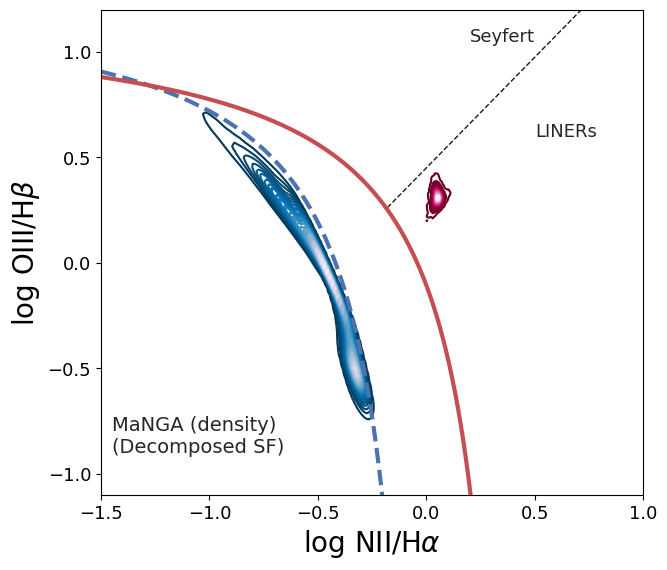}
    \includegraphics[width=5.8cm]{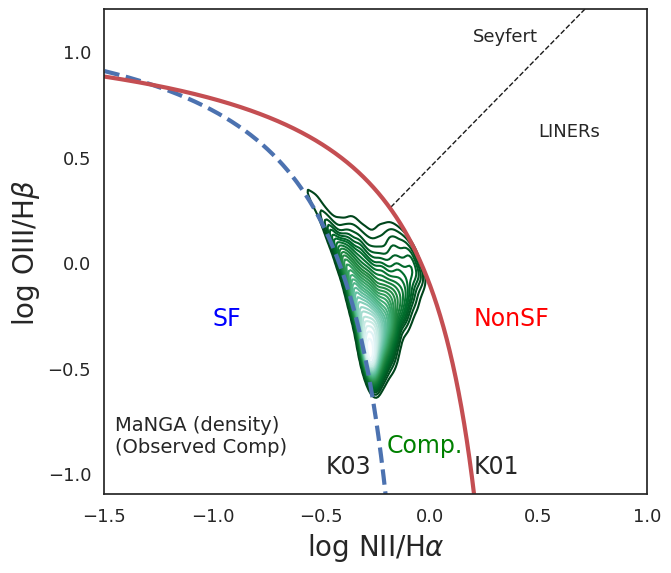}
    \includegraphics[width=6.6cm]{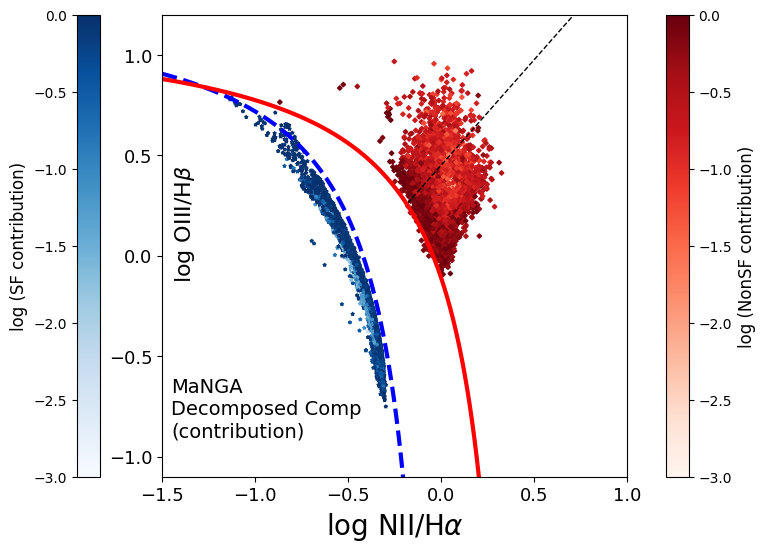}
    \includegraphics[width=5.4cm]{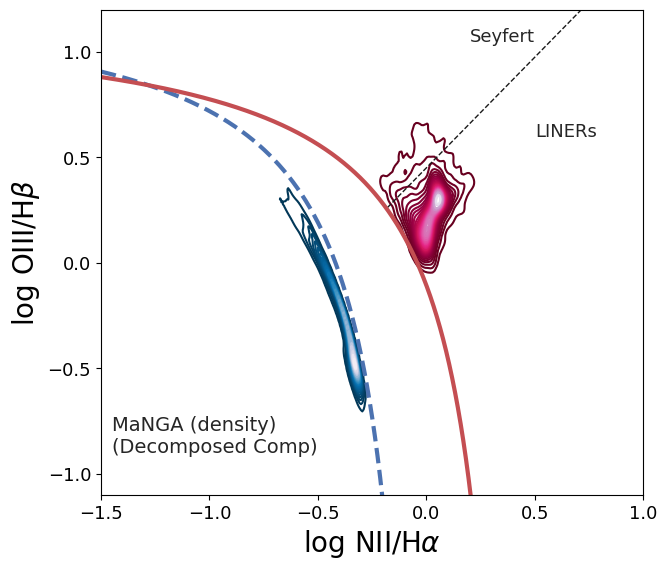}
    \includegraphics[width=5.8cm]{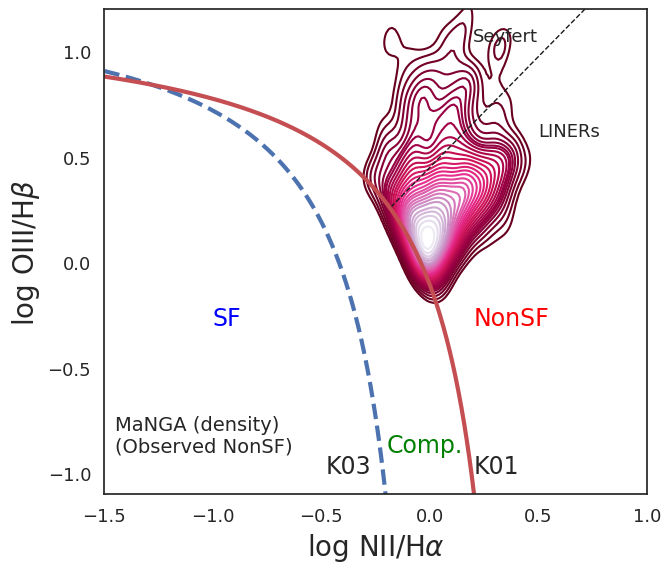}
    \includegraphics[width=6.6cm]{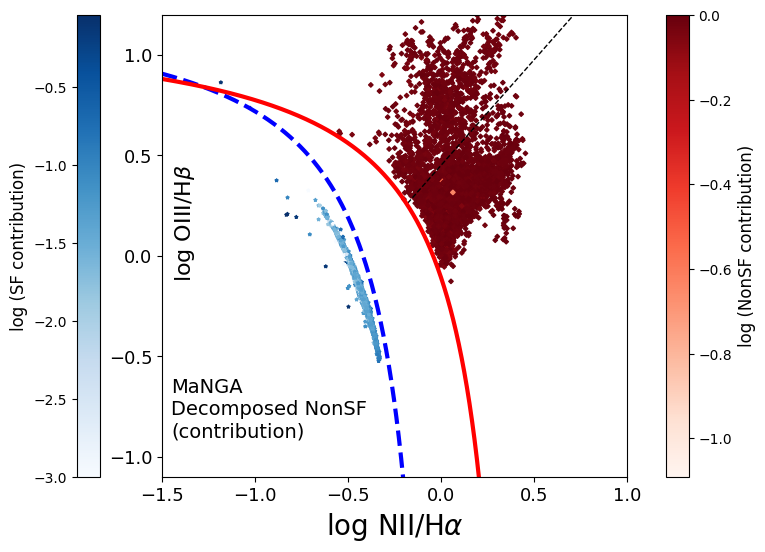}
    \includegraphics[width=5.4cm]{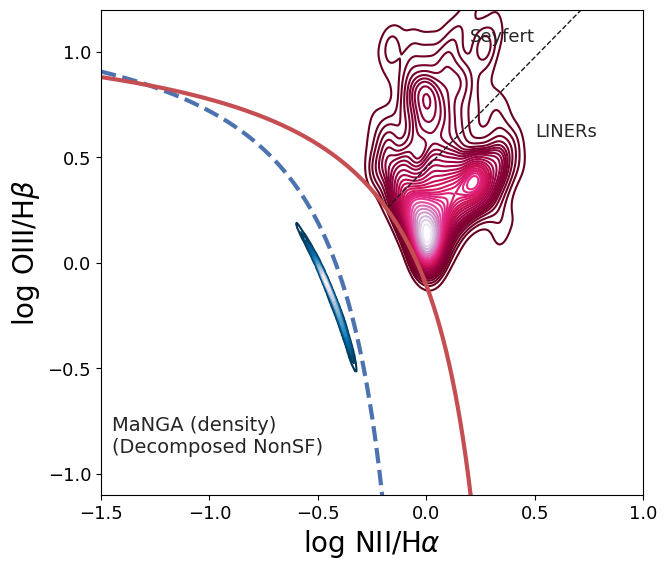}

    \caption{This plot is the same as Figure \ref{fig:fig-decompose-MaNGA}, but for the model that uses a continuum-subtracted training set. The two figures show very similar patterns.}
    \label{fig:fig-decompose-MaNGA-CS}
\end{figure*}

Figure \ref{fig:fig-decompose-MaNGA-CS} is similar to Figure \ref{fig:fig-decompose-MaNGA} (regarding the MaNGA data set), but for the model that utilizes a continuum-subtracted training set. The two figures exhibit very comparable patterns, validating that the network is indeed learning from the emission lines. A data-driven approach typically aims to minimize assumptions and utilize less theoretical models whenever possible. As demonstrated, the model can accurately predict the contributions to the emission lines regardless of the continuum. One advantage of using a data-driven approach (i.e., to use original spectra directly), for example, is that our DDM can detect continuum and bypass the additional step of continuum estimation.

\begin{figure*}[hbt!]
    \centering
    \includegraphics[width=5.6cm]{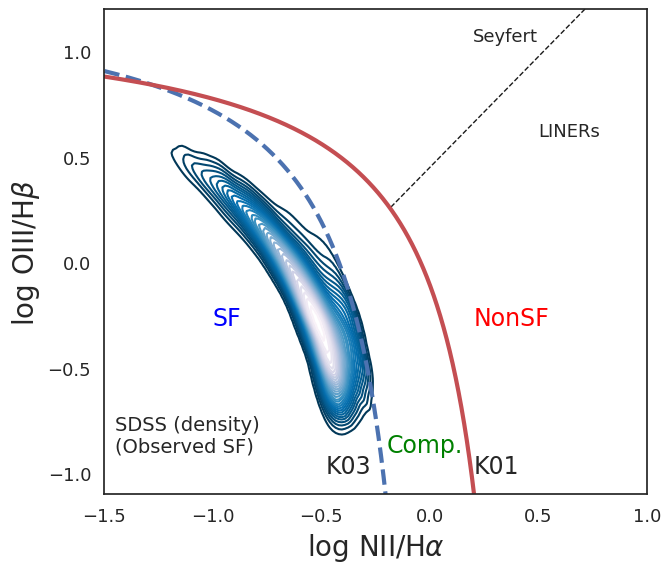}
    \includegraphics[width=6.6cm]{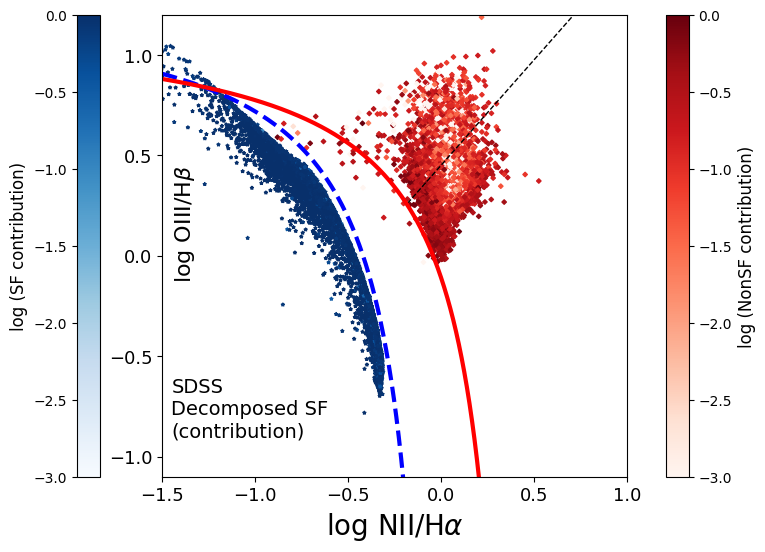}
    \includegraphics[width=5.5cm]{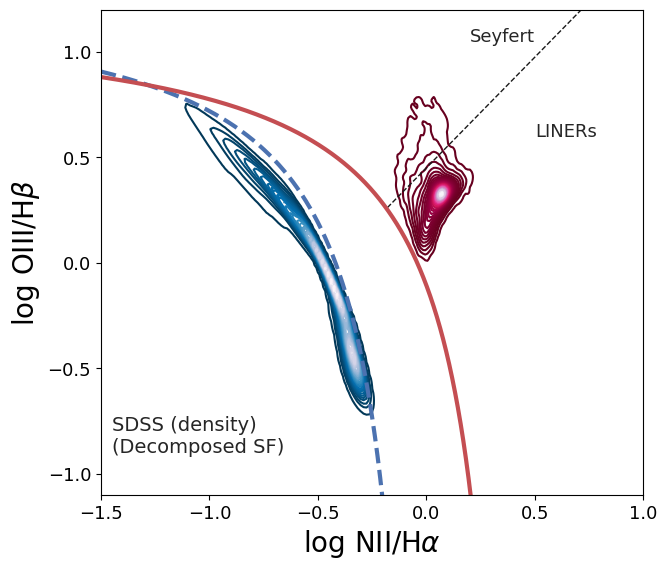}
    \includegraphics[width=5.6cm]{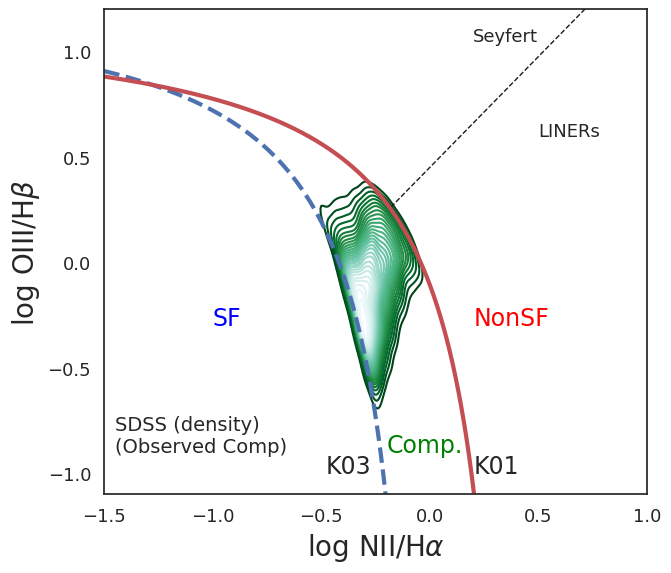}
    \includegraphics[width=6.6cm]{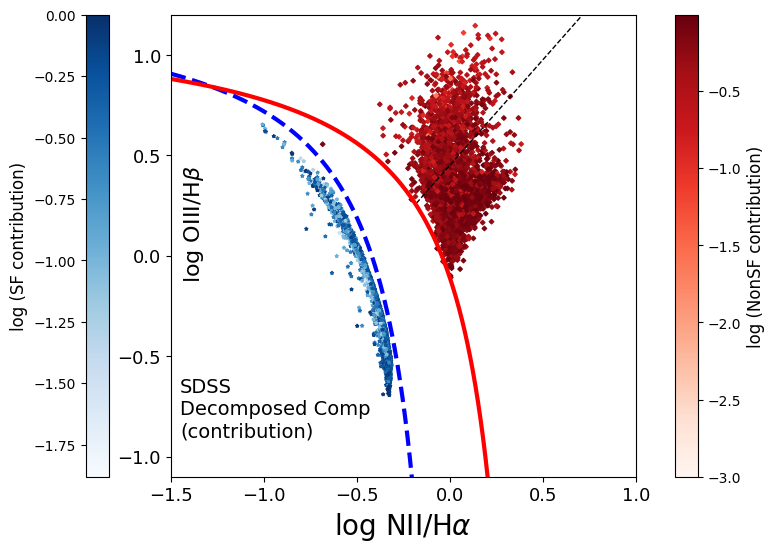}
    \includegraphics[width=5.5cm]{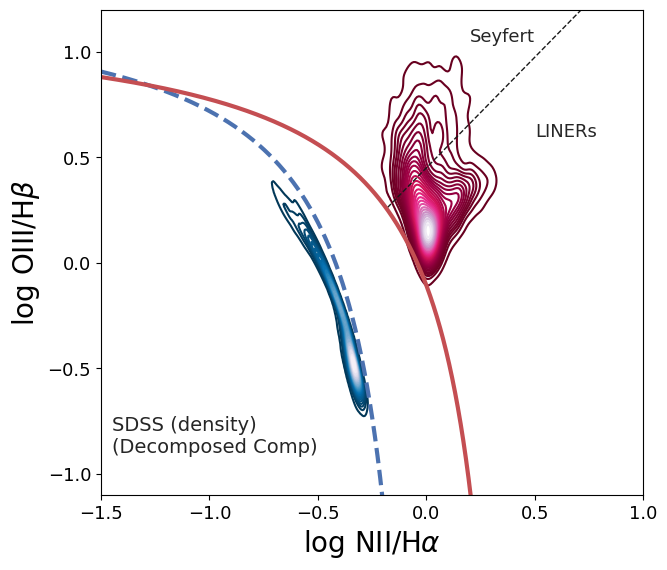}
    \includegraphics[width=5.6cm]{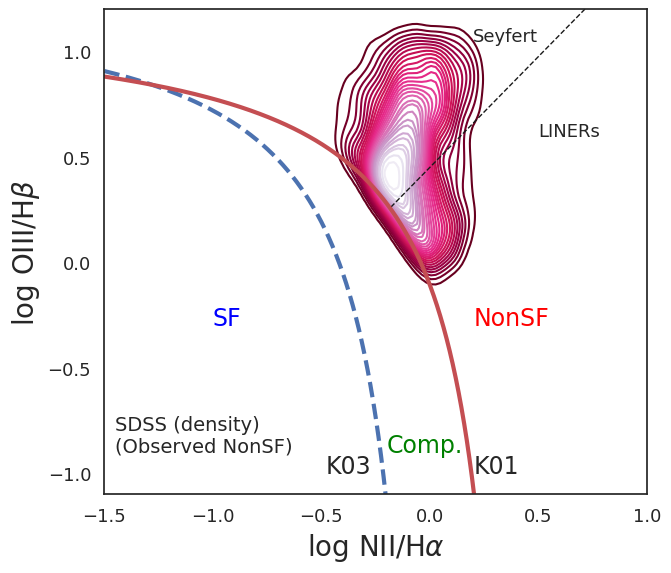}
    \includegraphics[width=6.6cm]{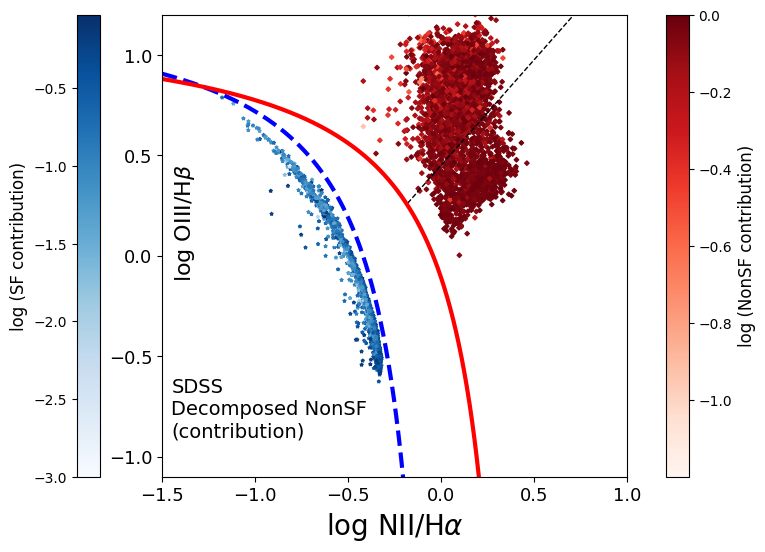}
    \includegraphics[width=5.5cm]{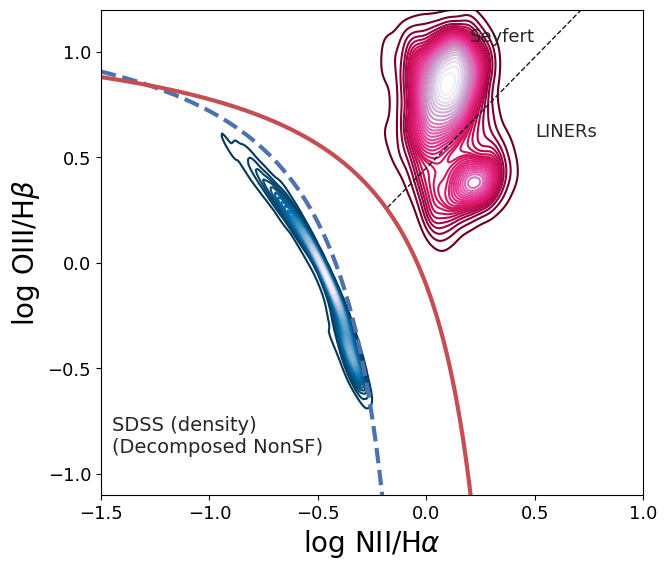}

    \caption{This plot is the same as \ref{fig:fig-decompose-MaNGA} but for the SDSS data set. The top left plot introduces SF spectra from SDSS processed through the model, revealing significant NonSF contamination, particularly in the LINER region, as shown in the top middle and right plots. This pattern reflects the mixed classes often resulting from single-fibre spectroscopy. In the middle panel, SDSS composite spectra are decomposed, demonstrating more extended populations and higher NonSF contributions in both SF and NonSF regions. The bottom panel illustrates the decomposition of NonSF spectra, highlighting a clear delineation into prominent clusters in the Seyfert and LINER regions after 'purification,' }
    \label{fig:fig-decompose-SDSS}
\end{figure*}

In Figure \ref{fig:fig-decompose-SDSS}, we present the same plot as in Figure \ref{fig:fig-decompose-MaNGA}, but it pertains to the SDSS dataset. Generally, we expect more 'contamination' and a different pattern here since we have a single-fibre dataset. In the top left plot of this figure, SF spectra from the SDSS dataset are fed into the model. The middle and right plots on the top display a more extended pattern in the NonSF region with a higher level of NonSF contribution (contamination) compared to the MaNGA SF spectra. This is not surprising because single-fibre spectroscopy typically results in more mixed classes. The `contaminated' part is mainly located in the LINER region of the NonSF. 

In the middle panel, the SDSS composite spectra are decomposed. Here, more extended populations with higher contributions (compared to SF) appear in both the SF and NonSF regions. The bottom panel, which displays the NonSF decomposition, shows that after 'purification,' the 'pure' NonSF spectra shift away from K01 (the red line) and clearly delineate two prominent clusters in the Seyfert and LINER regions. In other words, removing SF 'contamination' from SDSS NonSF spectra makes these two classes more distinct, although the contribution from the contamination part is not very high.

\begin{figure*}[hbt!]
    \centering
    \includegraphics[width=5.8cm]{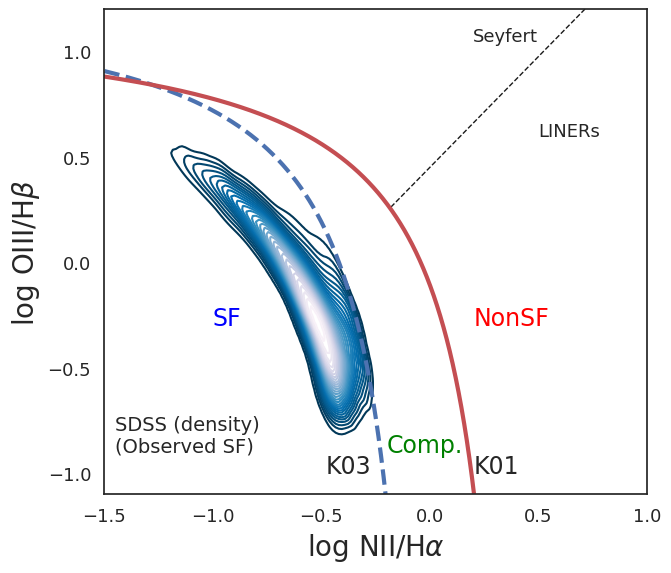}
    \includegraphics[width=6.6cm]{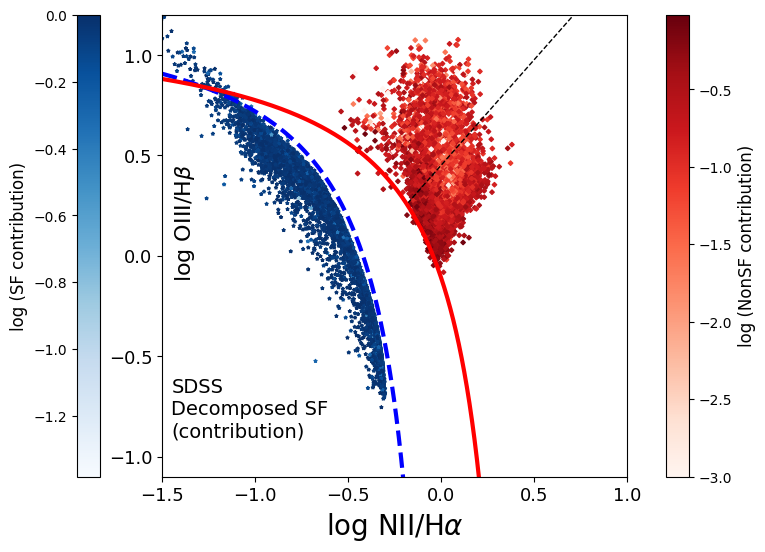}
    \includegraphics[width=5.4cm]{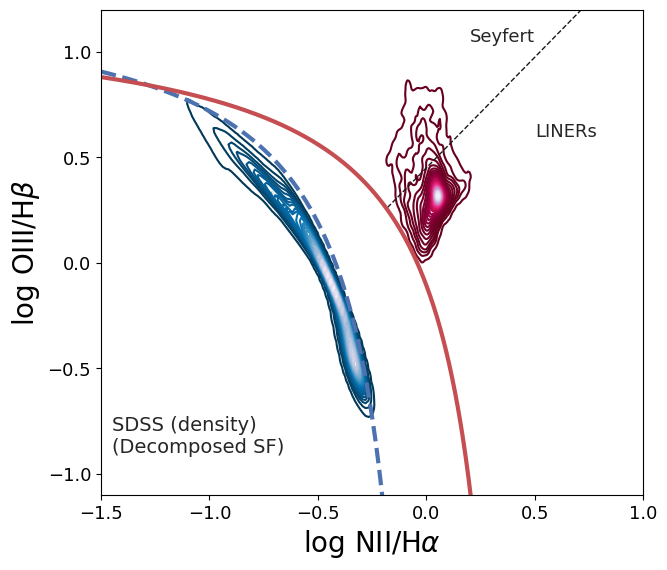}
    \includegraphics[width=5.8cm]{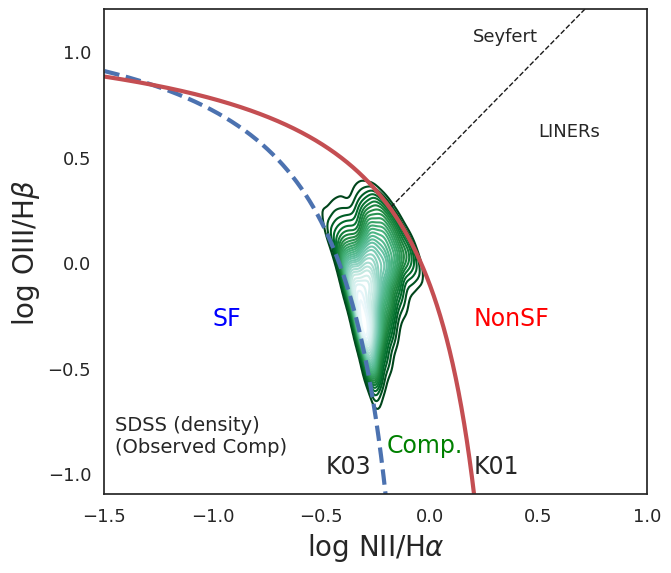}
    \includegraphics[width=6.6cm]{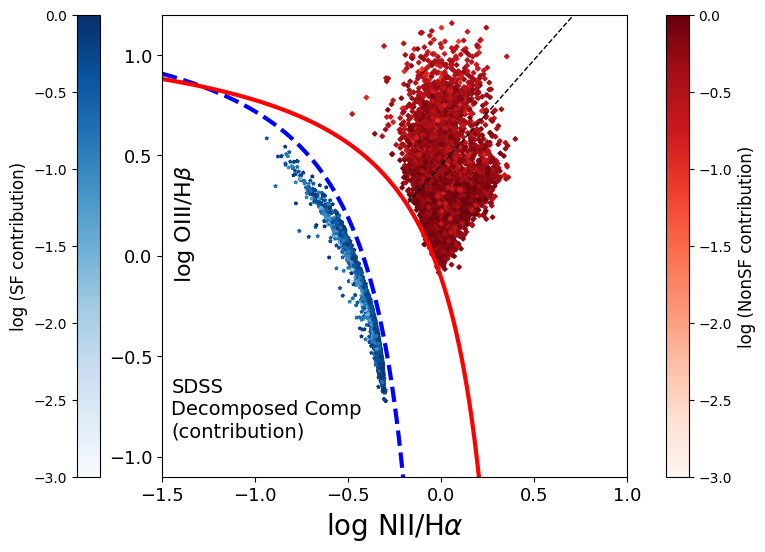}
    \includegraphics[width=5.4cm]{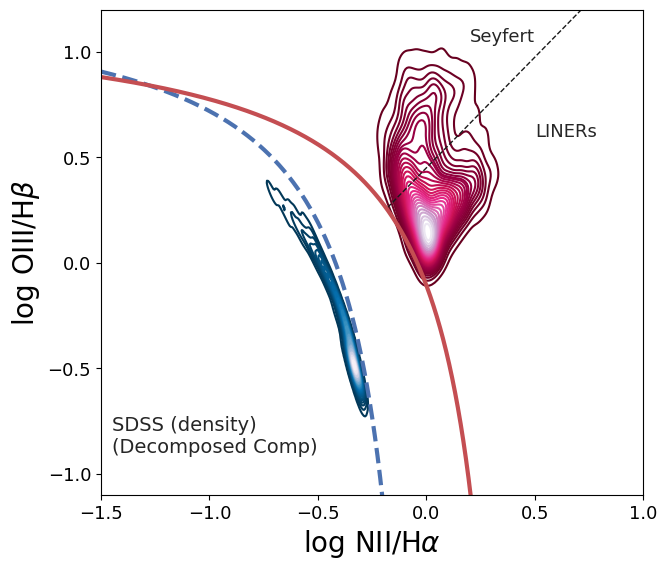}
    \includegraphics[width=5.8cm]{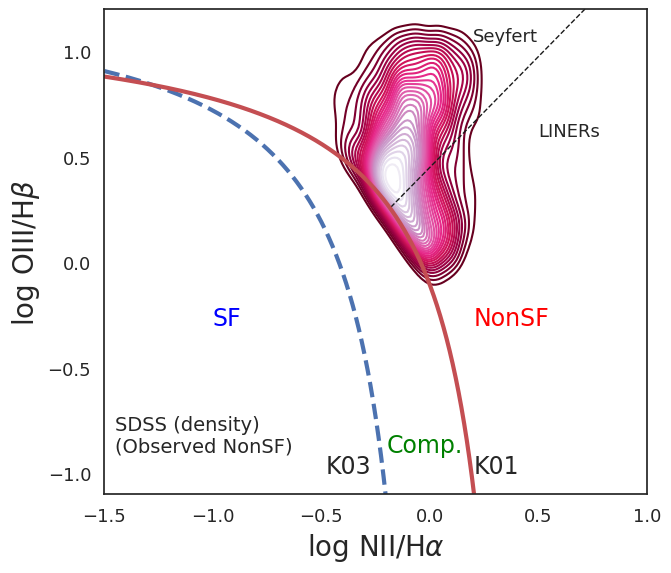}
    \includegraphics[width=6.6cm]{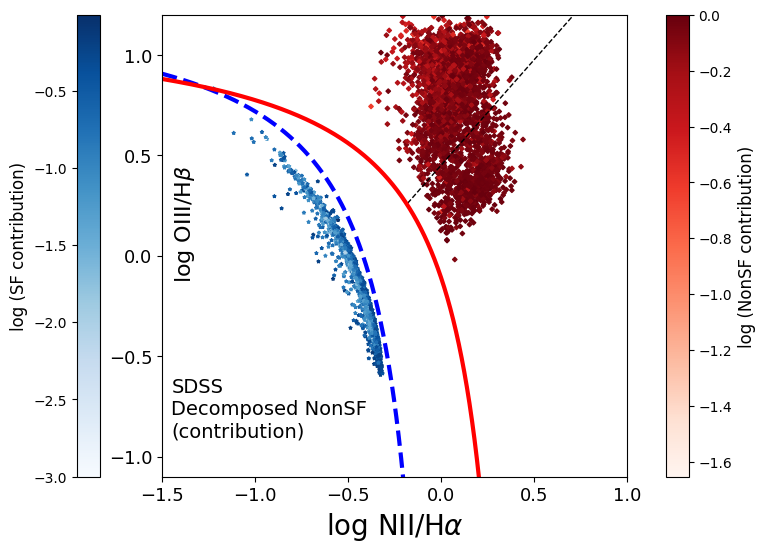}
    \includegraphics[width=5.4cm]{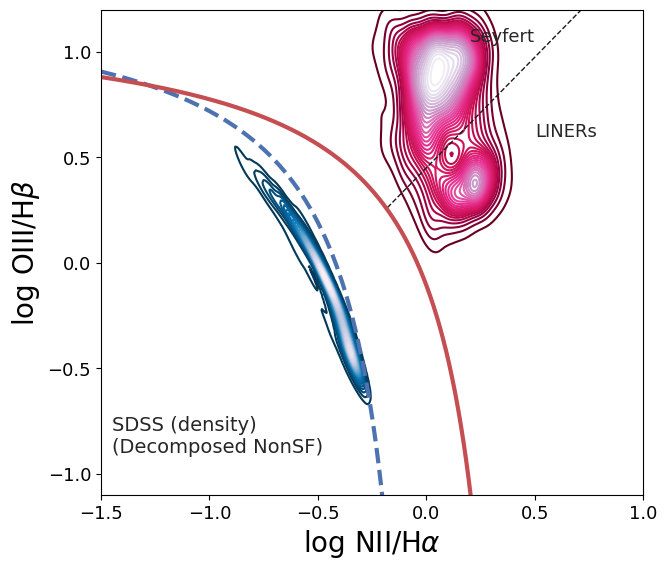}

    \caption{This plot is the same as Figure \ref{fig:fig-decompose-SDSS}, but for the model that uses a continuum-subtracted training set.}
    \label{fig:fig-decompose-SDSS-CS}
\end{figure*}

Figure \ref{fig:fig-decompose-SDSS-CS} resembles Figure \ref{fig:fig-decompose-SDSS}, but it pertains to the model using a continuum-subtracted training set. Both figures present similar patterns, confirming the consistency of the models.

\section{Summary and discussion}
\label{sec:summary}

We have devised a technique for decomposing a single spectrum by utilizing a training set and training various deep neural networks. This method employs the BPT diagram to classify spectra as either 'pure' SF (Star Forming) or NonSF (Non-Star Forming). To obtain 'pure' spectra for the training set, we utilized MaNGA spectra and trained three different deep neural networks: ratio-DNN, cont-DNN, and the main network, the Deep Decomposition Model (DDM). ratio-DNN, one of the networks, was trained to predict flux ratios for constructing BPT diagrams. We found that direct ratio estimations displayed high accuracy and did not exhibit systematic errors.

When decomposing a spectrum consisting of 2208 data points into two spectra, SF and NonSF, each also with 2208 data points, we considered two different scenarios. In one scenario, we input the spectrum in its original format, while in the other, we feed the model 2208 data points of the spectrum after subtracting the continuum. To subtract the continuum, we use our trained cont-DNN model. When we input an original spectrum into the model, the output is the best-fitted continuum to the spectra.

Once we have both the continuum and the original spectra, we can train two different DDMs to check whether the model is learning from the different stellar populations underlying the emission lines or from the emission lines themselves.  One DDM is for the original spectra, and another is for continuum-subtracted inputs. The decomposition results from both DDMs show very similar patterns on the BPT diagrams, indicating the model's capability to extract relevant information from the emission lines in the original spectra, not just the continuum. This demonstrates that with a large and informative training set, the model can account for the continuum in discriminating between SF and NonSF sources. This capability can reduce the number of preprocessing steps required to estimate the continuum in training sets.

To train a DDM, we need to synthesize 'combined' spectra from known 'pure' SF and NonSF sources, weighted by random numbers between 0 and 1, representing the SF contribution in the spectra. In addition to an independent validation set of synthesized spectra, we have also applied our method and networks to the MaNGA and SDSS datasets, as well as to an unusual SDSS spectrum, SDSS J1042-0018. For the MaNGA dataset, although we started with the assumption of 'pure' MaNGA SF and NonSF spectra, the results indicate that some purification is necessary. This is not surprising considering that, although the survey covers smaller regions than single-fibre spectroscopy, it can still encompass a blend of heterogeneous sources. However, this heterogeneous effect has a low impact, i.e., a low contribution of contamination for each spectrum in the training set.

The impact is more noticeable when we apply the DDMs to the SDSS spectra. As expected, there is more NonSF 'contamination' in SDSS SF galaxies due to the nature of the single-fibre survey, which is more likely to include mixed regions. An interesting case occurs when SDSS NonSF spectra are fed into the model. Here, the results reveal that once the SF 'contamination' is removed from the input spectra, a vertical distribution pattern emerges on the BPT diagram. This pattern contains two completely separated clusters in the NonSF region. The new distribution moves away from the line that separates NonSF spectra from the rest.  The two clusters of LINER and Seyfert regions are well separated, and this separation is in good agreement with the two LINER and Seyfert regions introduced by \cite{Kewley06}.

The spectrum of SDSS J1042-0018 displays distinct features that differ from those in our training set and do not resemble average SDSS spectra either. Estimating the fluxes, including flux ratio estimations, in this case, is challenging and often leads to uncertain results. However, our models can provide quick results that are reasonably close to expectations and consistent with other studies. Our method can be effectively used for data mining and deriving insights from a case based on a specific training set.

In this paper, we aim to demonstrate the method and the results based on a specific training set, along with the potential applications of our approach. One such application could be to estimate a more reliable SFR from sources potentially contaminated by other emissions. Here, we have presented our method as a two-class decomposition problem using the MaNGA survey. This means that any new input is projected onto the space created by DDM using the IFU survey.  An alternative training approach could expand this to a multi-component scenario. For instance, in the cases of availability, by using a large and proper training set of 'pure' SF, LINER, and Seyfert spectra, the decomposition could be expanded to distinguish between the three classes and their respective contributions. A multi-component scenario would be expected to show much better reconstructions. Another potential approach is to use theoretical training sets and compare the results with those from a data-driven approach.

In this paper, we have focused on the method and presented various results obtained from it. We have not yet followed up on extracting the physical parameters from the decomposed spectra, with the exception of a few flux ratio estimations. This process requires fitting procedures using relevant software packages like pPFX, which can help find more connections between the combined spectra and their components. These connections might provide better insight into and interpretation of the decomposition results. Finally, we utilized the MaNGA data release 15, which was sufficient for our approach, although a newer release with an increased number of spectra is available.

\clearpage
\section*{Acknowledgement}
HT acknowledges support from an NSERC Discovery Grant. This research used the facilities of the Canadian Astronomy Data Centre operated by the National Research Council of Canada with the support of the Canadian Space Agency.

\bibliography{main,references}{}
\bibliographystyle{aasjournal}


\end{document}